\documentclass[structabstract]{aa}  
\usepackage{amstext,amsmath,mathtools}
\usepackage{graphicx}
\usepackage{subfigure}
\usepackage{textcomp}
\usepackage{txfonts}
\usepackage{natbib}
\usepackage{time}
\usepackage{subfig}
\usepackage{color}
\usepackage{url}
\bibpunct{(}{)}{;}{a}{}{,}
\usepackage{overpic}
\usepackage{fancyvrb}
\bibpunct{(}{)}{;}{a}{}{,}
\def\Blos{\ensuremath{B_\text{LOS}}}
\def\vlos{\ensuremath{v_\text{LOS}}}
\def\Ic{\ensuremath{I_\text{c}}}

\def\Imin{\ensuremath{I_\text{min}}}
\def\mps{m\,s$^{-1}$}
\def\kmps{km\,s$^{-1}$}
\def\figwidth{0.91\textwidth}

\begin{document}

\title{Transient downflows associated with the intensification of small-scale magnetic features and bright point formation}

\author{G. Narayan\inst{1,2}}

\institute{Institute for Solar Physics, Royal Swedish Academy of
  Sciences, AlbaNova University Center, 106\,91 Stockholm, Sweden
  \\
  \and Stockholm Observatory,
  Dept. of Astronomy, Stockholm University,
  AlbaNova University Center, 106\,91 Stockholm, Sweden\\
\email{gautam@astro.su.se}
} \date{Draft: \now \today}

\abstract
{Small-scale magnetic features are present everywhere in the solar photosphere. Theoretical models, numerical calculations, and simulations describing the formation of these features have existed for a few decades, but there are only a few observational studies in direct support of the simulations. In particular, there is a lack of high resolution spectropolarimetric observations capturing the formation of small-scale magnetic features.}
{In this study we present the evolution of small-scale magnetic features with a spatial resolution close to 0\farcs15 and compare these observations with those predicted by numerical simulations and also with previous observational work of a similar nature.}
{We analyze a 40 min time sequence of full Stokes spectropolarimetric 630.25~nm data from a plage region near the Sun center. We use line-of-sight velocities and magnetic field measurements obtained using Milne-Eddington inversion techniques with and without stray-light compensation along with measured continuum and line minimum intensities. We discuss the results in relation to earlier observations and simulations}
{We present eight cases involving strong \emph{transient} downflows and magnetic field intensification. All cases studied are associated with the formation of a bright point in the continuum. In three out of the eight cases we find the presence of weak opposite polarity field in close proximity to the downflow.}
{Our data are consistent with earlier simulations describing flux tube collapse, but the transition to a state with stronger field appears transient and short-lived, rather than resulting in a permanent field intensification. Three cases of weak opposite polarity field found adjacent to the downflows do not appear related to reconnection but may be related to overturning convection pulling down some field lines and leading to up/down "serpentine" field, as seen in some simulations.}

\keywords{Sun: photosphere -- Sun: granulation -- Sun: faculae, plages -- Sun: surface magnetism}
\titlerunning{Transient downflows associated with magnetic field intensification}
   \maketitle

\section{Introduction}
\label{sec:introduction}
The solar photosphere is permeated by magnetic field manifesting itself in various small-scale features. The formation of small-scale magnetic features is explained by convection in which the magnetic field lines embedded in the solar plasma is concentrated into intergranular lanes by the horizontal component of granular convection. This concentrated flux is often referred to as a flux tube. The flux tube prevents heat transfer and the gas inside the flux tube is isolated leading to radiative cooling. This triggers a convective instability in the form of an accelerated downflow resulting in partial evacuation of the flux tube, and further intensification of its field strength. This process is referred to as convective collapse \citep{1978ApJ...221..368P,1979SoPh...61..363S}. The observational signature of such a process is the presence of strong downflows within the fluxtube accompanying the intensification of its magnetic field strength. Partial evacuation also causes a depression of the $\tau=1$ surface, where $\tau$ is the optical depth, and since the flux tube is hotter deeper down, it appears brighter.

The convective collapse process is limited to kilo-gauss (kG) strength flux tubes in the solar photosphere. The explanation for the existence of sub-kG flux tubes has been attributed to the efficient radiative heat exchange with the surrounding gas for the smaller flux tubes \citep{1986Natur.322..156V}, preventing complete evacuation of the flux tube. 2-D MHD simulations of \citet{1998A&A...337..928G}  have shown that convective collapse can lead to the formation of both stable and unstable kG configurations.

Until recently, studies relating to the formation of small-scale magnetic structures have mostly been theoretical, with only very little observational evidence of convective collapse. This is due to the small spatial scales involving such events. Some of the earliest examples of possible convective collapse observations were reported by \citet{1999ApJ...514..448L}, \citet{1999A&A...349..941S}, and \citet{2001ApJ...560.1010B}. More recent observational evidence for convective collapse has been provided by \citet{2008ApJ...677L.145N}, \citet{2008ApJ...680.1467S}, and \citet{2009A&A...504..583F}, using Hinode data, and \citet{2009A&A...494.1091B} using G\"ottingen FPI data. \citet{2010A&A...509A..76D} compared 3-D MHD simulations of magnetic field intensification with simulated Hinode observations of the same events.

Downflows have also been found adjacent to magnetic field concentrations. This is attributed to the cooling of the surrounding gas, which leaks radiation through the optically thin flux tube \citep{1984A&A...139..435D}. Downflows at the edge of bright magnetic features were reported by \citet{2004ApJ...604..906R} and close to pores \citep{2001ApJ...552..354L,2003ApJ...598..689S,2003A&A...405..331H,2005A&A...432..319S}. Downflows occurring at the boundary of flux concentrations have also been seen in magneto-convection simulations of \citet{2004RvMA...17...69V} and  surrounding pores in the simulations of \citet{2007A&A...474..261C}. In a recent paper \citep{2010arXiv1007.4673N} it was found that almost all the downflows associated with \emph{bright} magnetic features in strong plage occur at the \emph{boundaries} of the strong field.

Downflows have also been observed in association with small-scale flux emergence in a plage region \citep{2008A&A...481L..25I}. As the flux tubes rise, they carry material with them; this rising material eventually must flow down. \citet{2008ApJ...687.1373C} obtained such downflows in their simulations of emerging flux. They also found supersonic downflows associated with flux-cancellation events, which took place when flux concentrations of opposite polarity within an emerging flux region came in contact. They identify the Lorentz force as responsible for these supersonic downflows.

Downflows were also found in association with sunspots. \citet{2008ApJ...680.1467S} demonstrated the presence of high speed downflows related to the formation of moving magnetic features in the outer boundary of the penumbra in a well developed sunspot. The same authors also found strong downflows at the edge of the part of the umbra of a sunspot which lacked penumbral structure. They explain this as due to the interaction of convection with a strong magnetic field similar to that of downflows in the periphery of pores.

In this paper we present and analyze spectropolarimetric observations at a resolution close to 0\farcs15 that show transient downflows associated with the formation of small-scale bright magnetic features and, in three cases, with weak opposite polarity field.

\section{Observations and data processing}
\label{sec:observ-data-reduct}
The observations presented in this paper comprise a plage region close to disk center (heliocentric coordinates S9 E11) which was mostly unipolar within our field of view, as shown in Fig. 1. The data were recorded on 26 May 2008 with the CRisp Imaging SpectroPolarimeter \citep[CRISP; ][]{scharmer08crisp} at the Swedish 1-m Solar Telescope \citep[SST; ][]{2003SPIE.4853..341S}. A time series of approximately 40 minutes was obtained under variable seeing conditions. Full Stokes data were recorded at 9 wavelength positions spanning $-19.2$ to $+19.2$~pm from the line center of the \ion{Fe}{i} 630.25~nm in steps of 4.8~pm and at a continuum wavelength. The data were processed using the MOMFBD image reconstruction technique \citep{lofdahl02multi-frame,noort05solar}, demodulated, and then corrected for telescope polarization. The final Stokes data was obtained after correcting the demodulated data for residual $I$ to $Q$, $U$, and $V$ cross-talk using the Stokes data recorded in the continuum. The polarimetric noise levels were estimated to be $1.2\times 10^{-3}$, $1.4\times 10^{-3}$, and $1.0\times 10^{-3}$ for Stokes $Q$, $U$, and $V$, respectively.

We ran Milne-Eddington inversions on the reduced data using the Helix code \citep{2004A&A...414.1109L}. This code computes synthetic Stokes profiles based on Unno-Rachkovsky solutions of the polarized radiative transfer equations. The synthesized profiles are convolved with the CRISP filter profile and fitted with the observed profiles using the genetic algorithm Pikaia \citep{1995ApJS..101..309C}. The inversions were initially made assuming unit magnetic filling factor and zero stray-light. As discussed in Sect. 3, we also made inversions with $40\%$ assumed stray-light. From the inversions we obtained the magnetic field strength ($B$), line-of-sight (LOS) velocity (\vlos) and the inclination angle $\gamma$. The LOS magnetic field ($\Blos$) is computed as $B$ $\cos$ $\gamma$. In order to remove the Doppler-shift bias introduced in the data due to cavity errors and the location of the \`{e}talons close to the focal plane \citep{scharmer06comments}, the LOS velocity maps obtained from inversions were subtracted with a 6302 \ion{Fe}{i} line \`{e}talon cavity map that was made using flat-field images, obtained 1 h after recording the science data. We computed the \emph{average} pore Doppler shifts over the time series and used this value to set the zero point of \vlos\ maps. The absence of significant drifts was confirmed by plotting \vlos\ averaged over the center 650$\times$650 pixels. This showed small variations, with an amplitude close to 100 \mps and a period around 300 s, but no signs of any systematic drifts over time. We also compiled maps of the 630.25~nm line minimum intensity to help identify brightenings that occur in the continuum maps due to the higher contrast in the line-minimum maps.

The details of the observational setup, data processing, and inversions are described by \citet{2010arXiv1007.4673N}, who discuss a single snapshot from the data set discussed in the present paper.

\begin{figure*}[!htb]
  \begin{center}
    \includegraphics[bb=49 22 483 457, clip,height=0.45\linewidth]{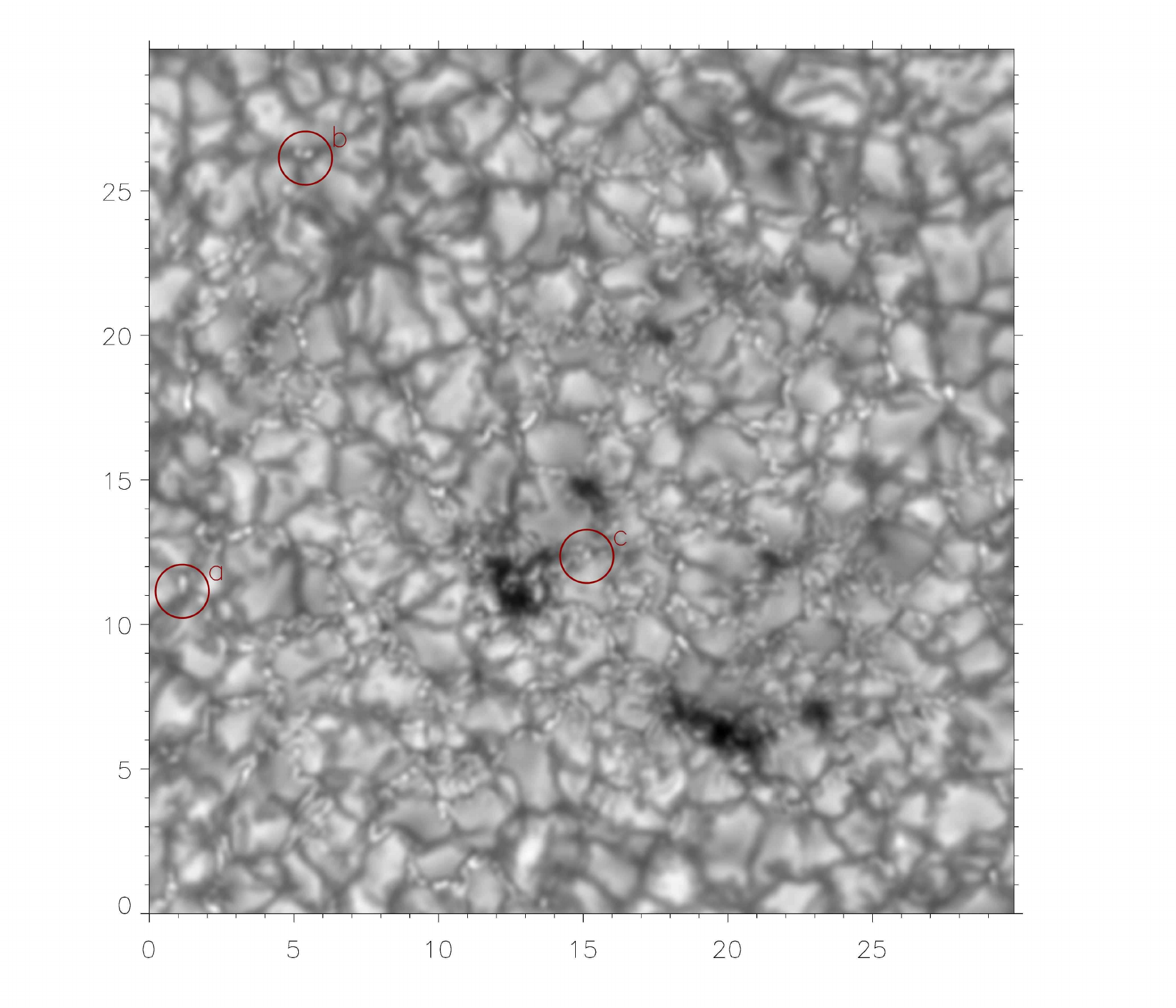}
    \includegraphics[bb=49 22 535 422, clip,height=0.45\linewidth]{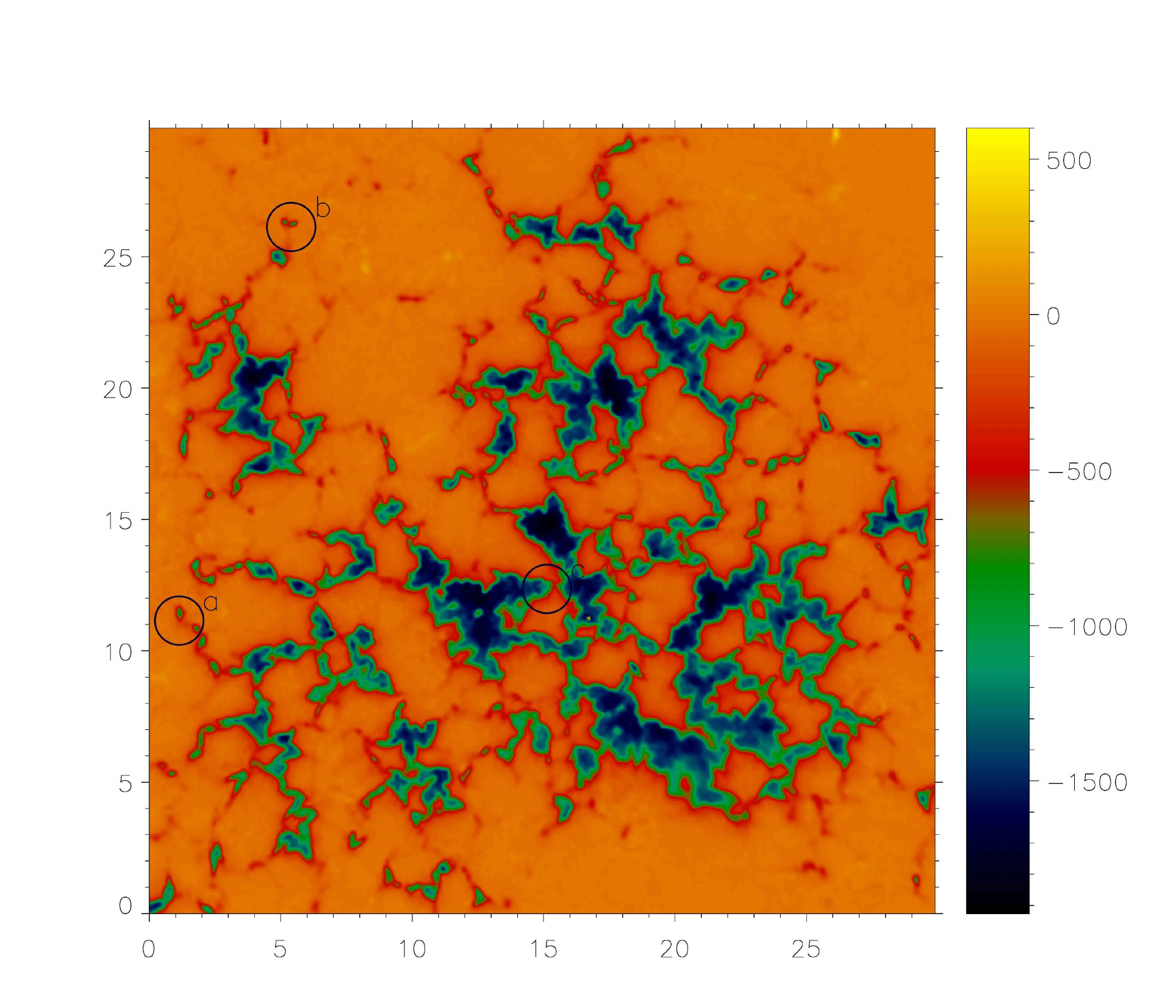}
     \caption{A part of the field of view showing the nearly unipolar region with small pores. The left panel shows the restored CRISP
      continuum image, the right panel shows the LOS magnetic flux density, $\Blos$, obtained from the inversions assuming $40\%$ stray-light. The circles in both maps show the locations of the first three downflow cases discussed in the paper. Tick marks are in units of arcseconds. The color bar indicates the signed LOS magnetic field in gauss.}
    \label{fig:fullFOV}
  \end{center}
\end{figure*}

\renewcommand{\thesubfigure}{{}}
\begin{figure*}[htb!]
  \subfigure[ Case-a]{
   \includegraphics[bb=45 95 1133 640,clip,width=0.240\linewidth]{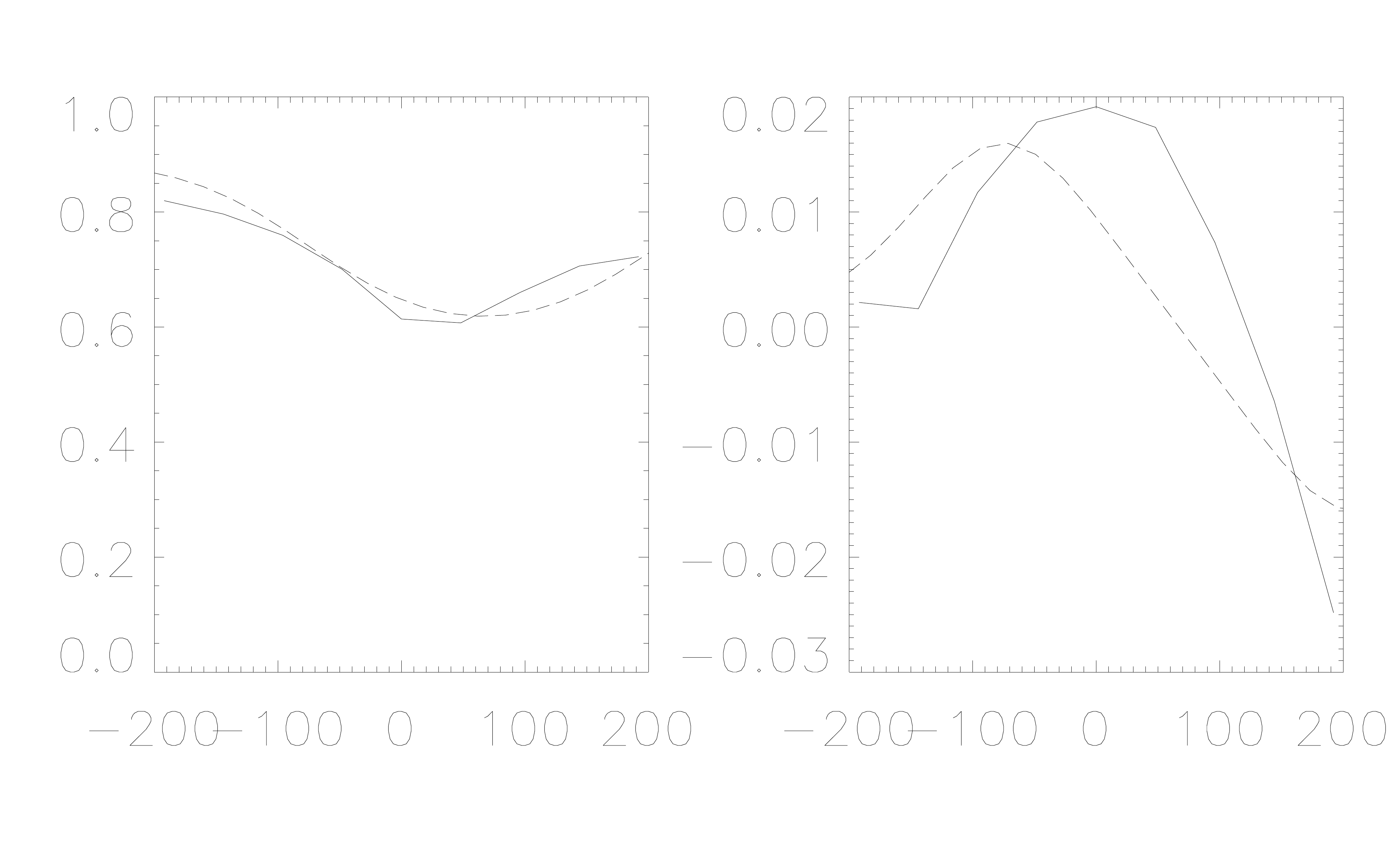}}
  \subfigure[ Case-b]{
    \includegraphics[bb=45 95 1133 640,clip,width=0.240\linewidth]{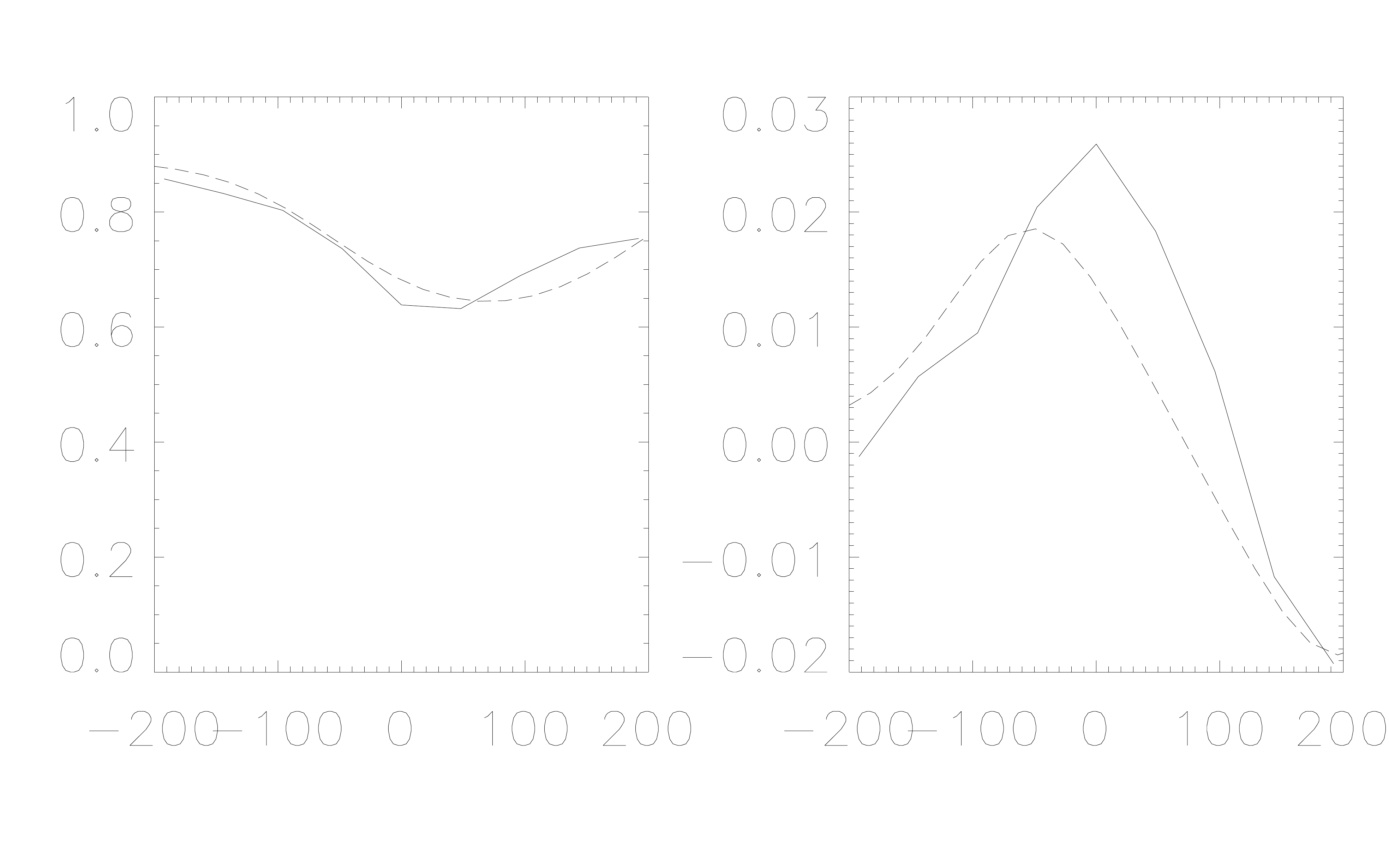}}
  \subfigure[ Case-c]{
    \includegraphics[bb=45 95 1133 640,clip,width=0.240\linewidth]{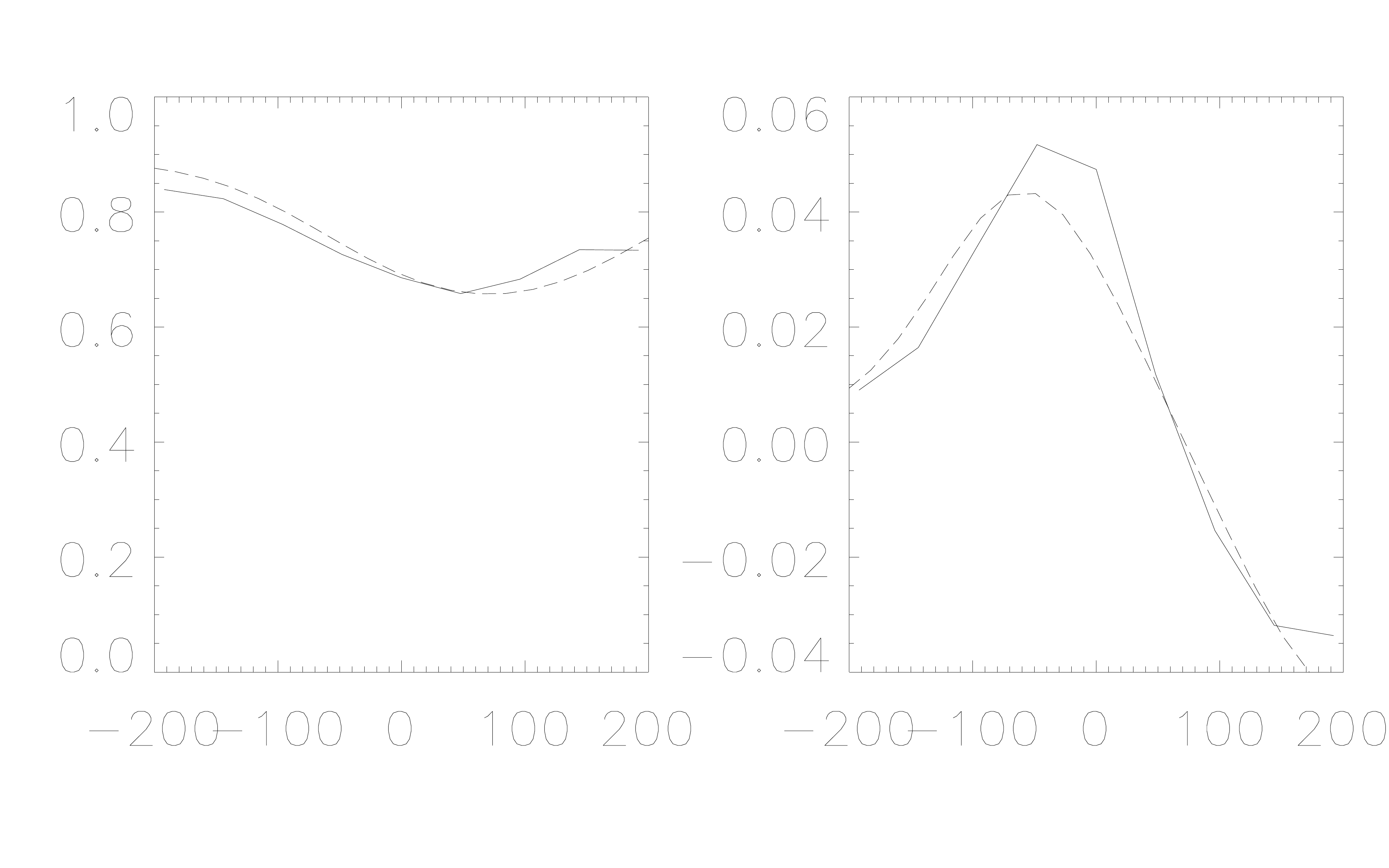}}
   \subfigure[ Case-d]{
    \includegraphics[bb=45 95 1133 640,clip,width=0.240\linewidth]{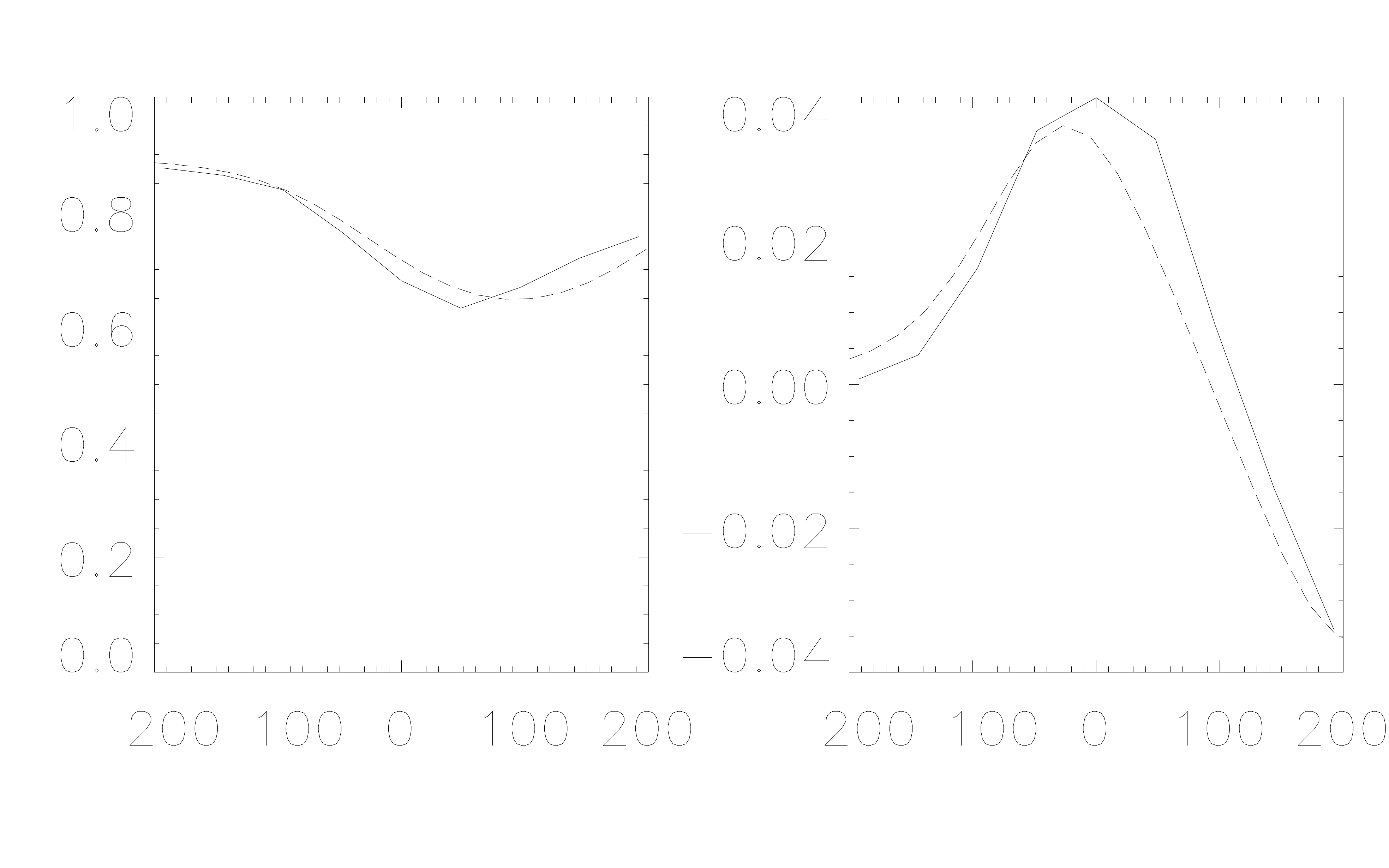}}
   \subfigure[ Case-e]{
    \includegraphics[bb=45 95 1133 640,clip,width=0.240\linewidth]{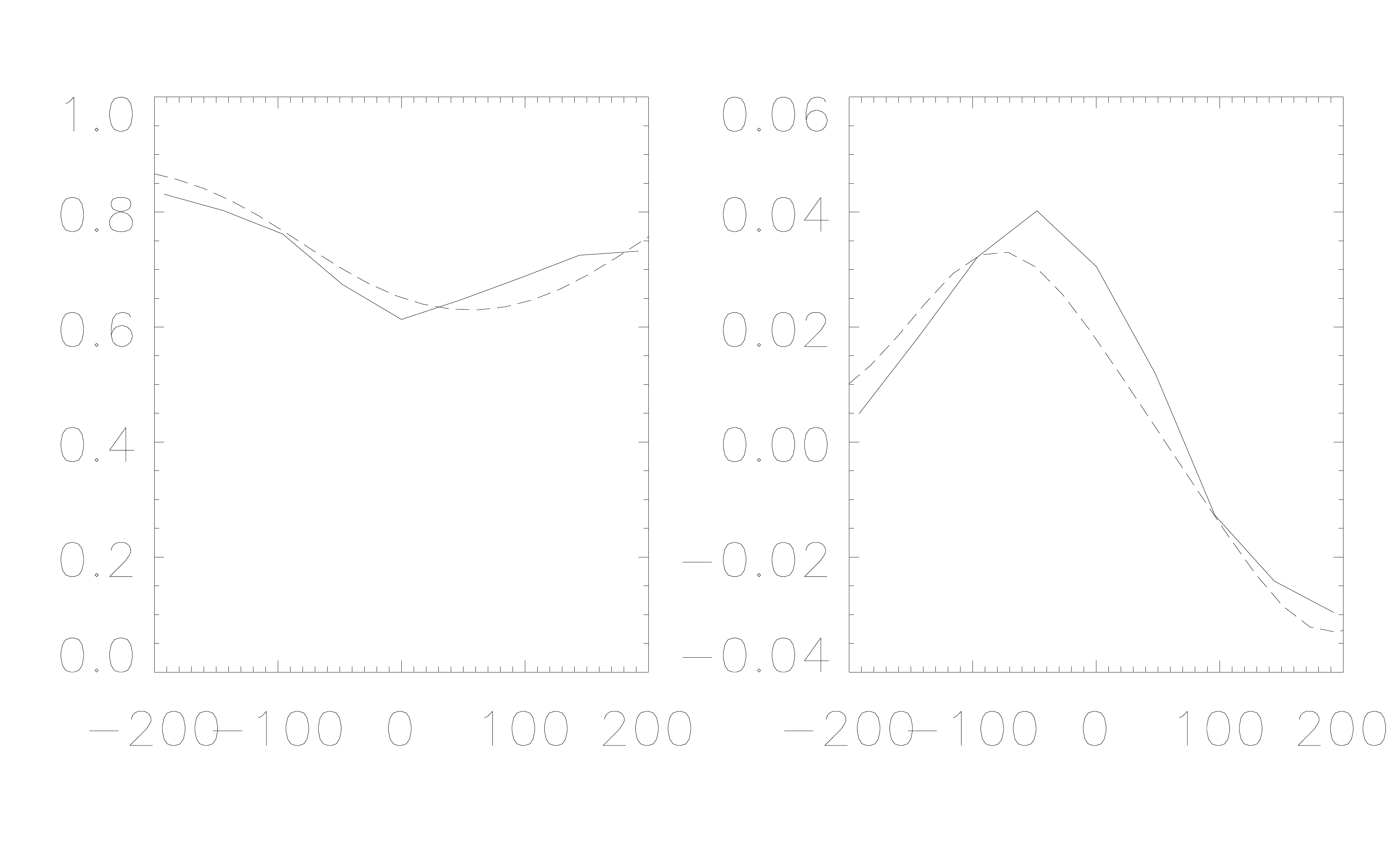}}
   \subfigure[ Case-f]{
    \includegraphics[bb=45 95 1133 640,clip,width=0.240\linewidth]{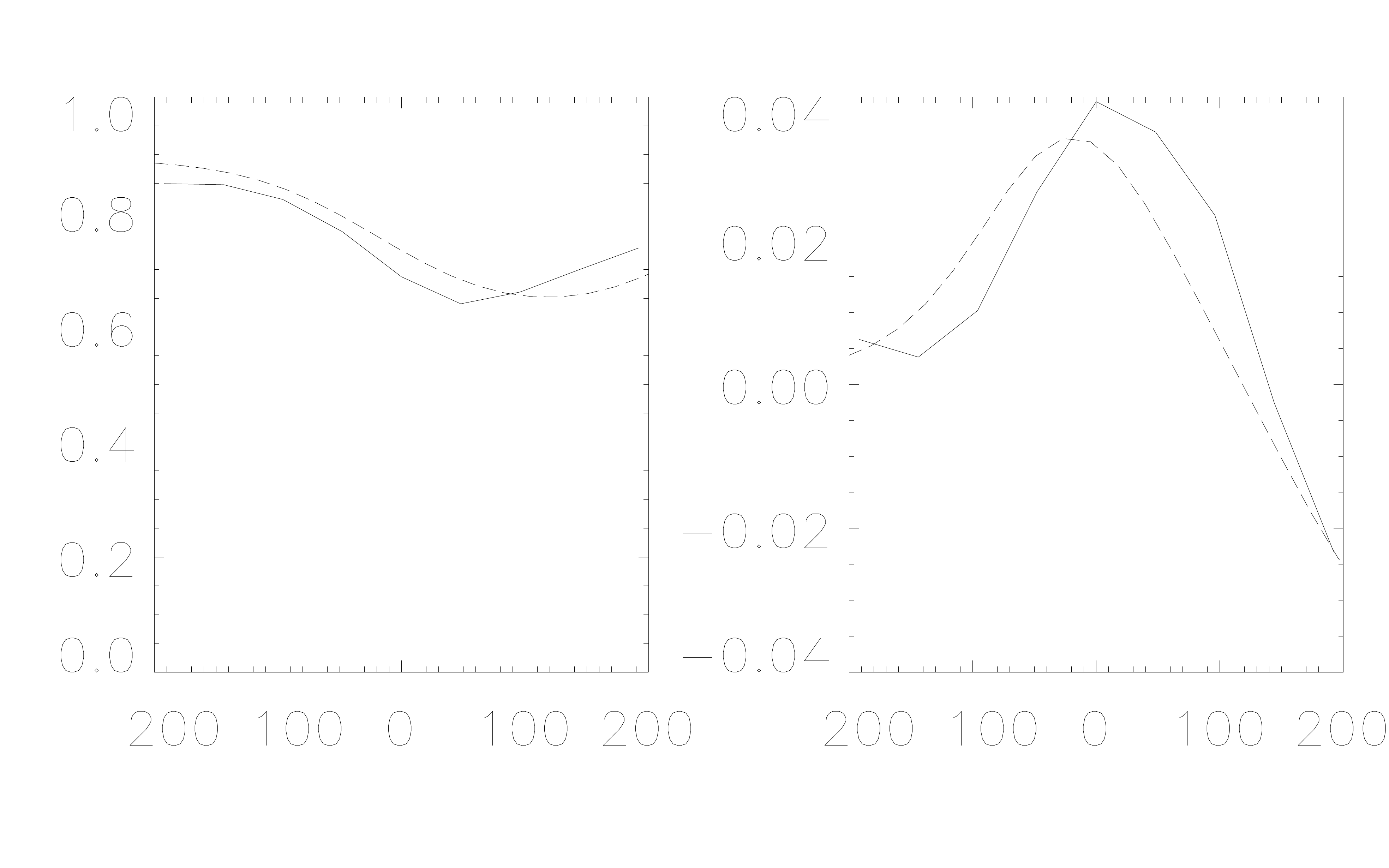}}
    \subfigure[ Case-g]{
    \includegraphics[bb=45 95 1133 640,clip,width=0.240\linewidth]{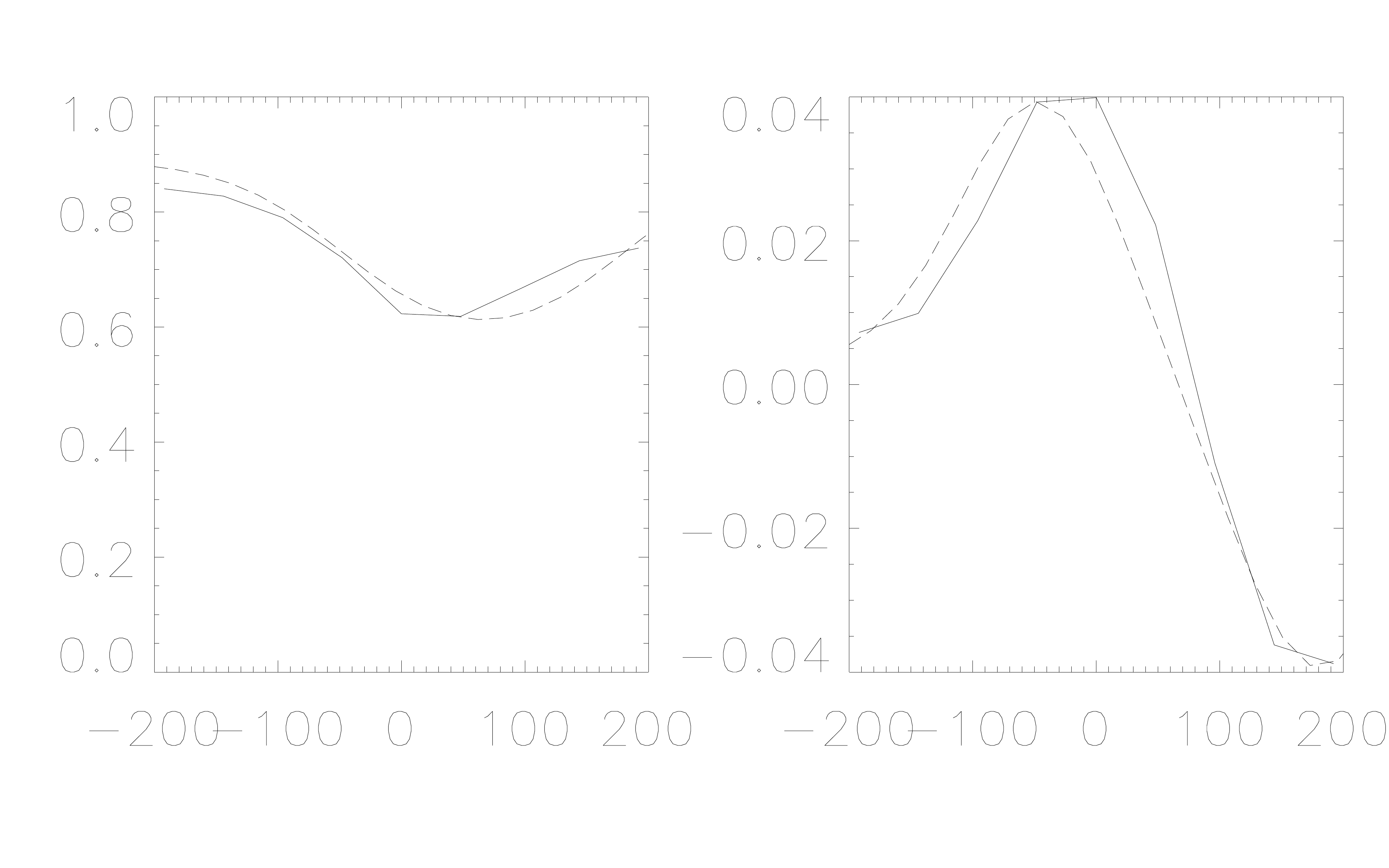}}
  \subfigure[ Case-h]{
    \includegraphics[bb=45 95 1133 640,clip,width=0.240\linewidth]{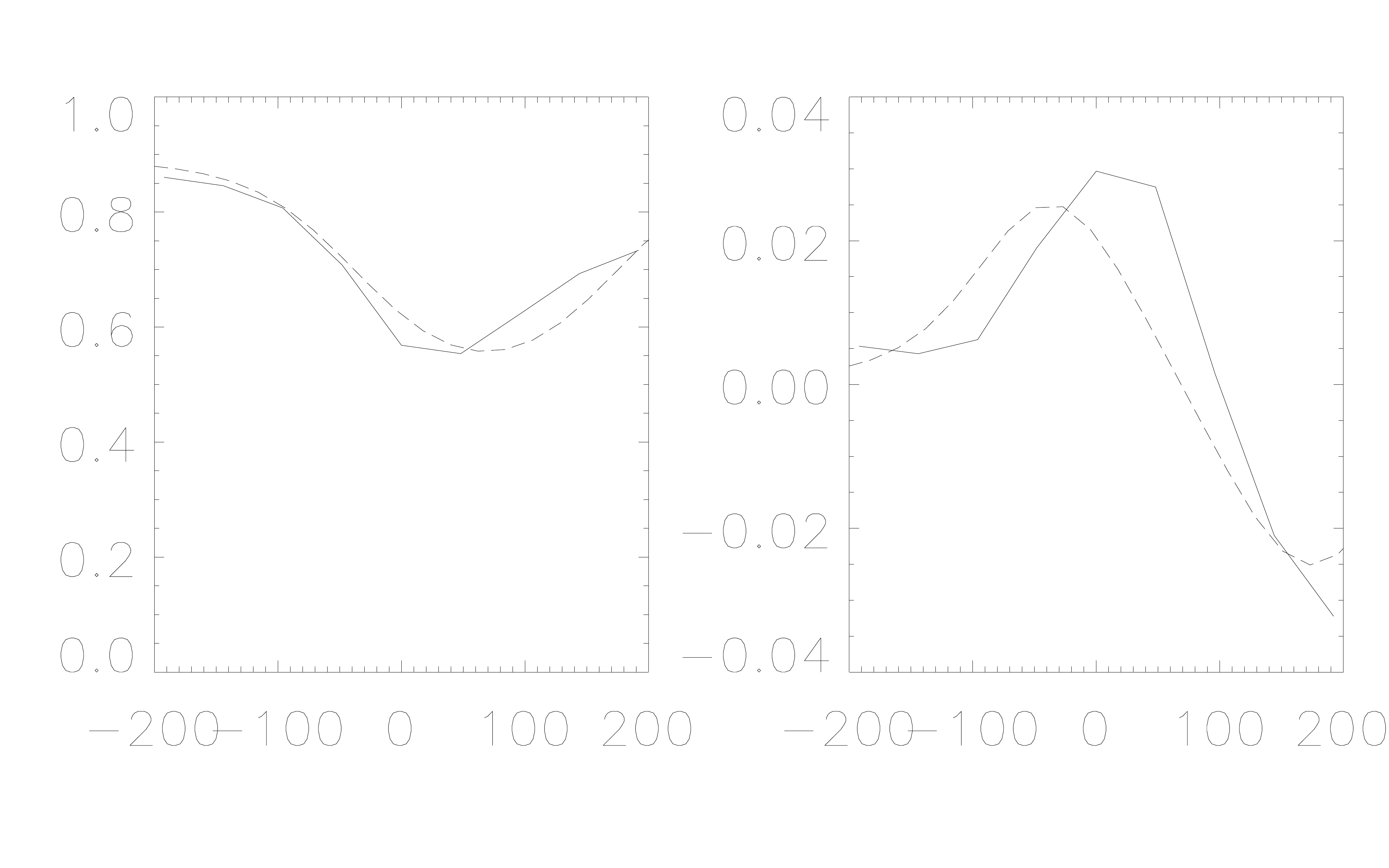}}
   \caption{Observed (solid) and fitted (dashed) Stokes $I$ and $V$
    profiles corresponding to the peak \vlos\ in the different cases without stray-light. Both Stokes $I$ and $V$ are normalized to the continuum intensity, the wavelength of which is outside the range plotted. Wavelengths along the x-axis are in units of m{\AA}.}
  \label{fig:fitobs}
\end{figure*}
\begin{figure*}[!htbp]
       \subfigure[Case-a]{
    \includegraphics[bb=55 14 450 323, clip,height=0.18\linewidth]{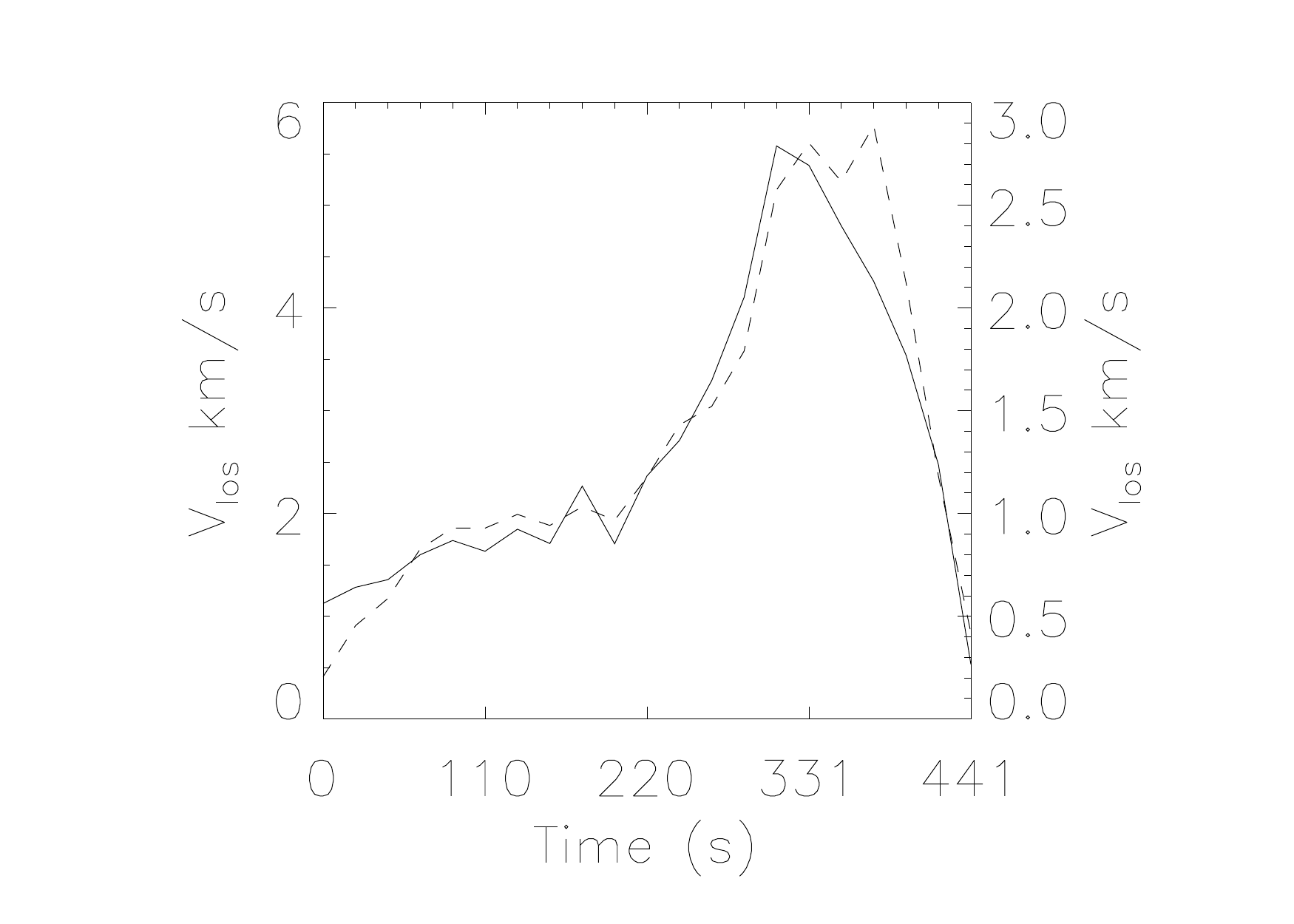}}
\subfigure[Case-b]{
    \includegraphics[bb=55 14 450 323, clip,height=0.18\linewidth]{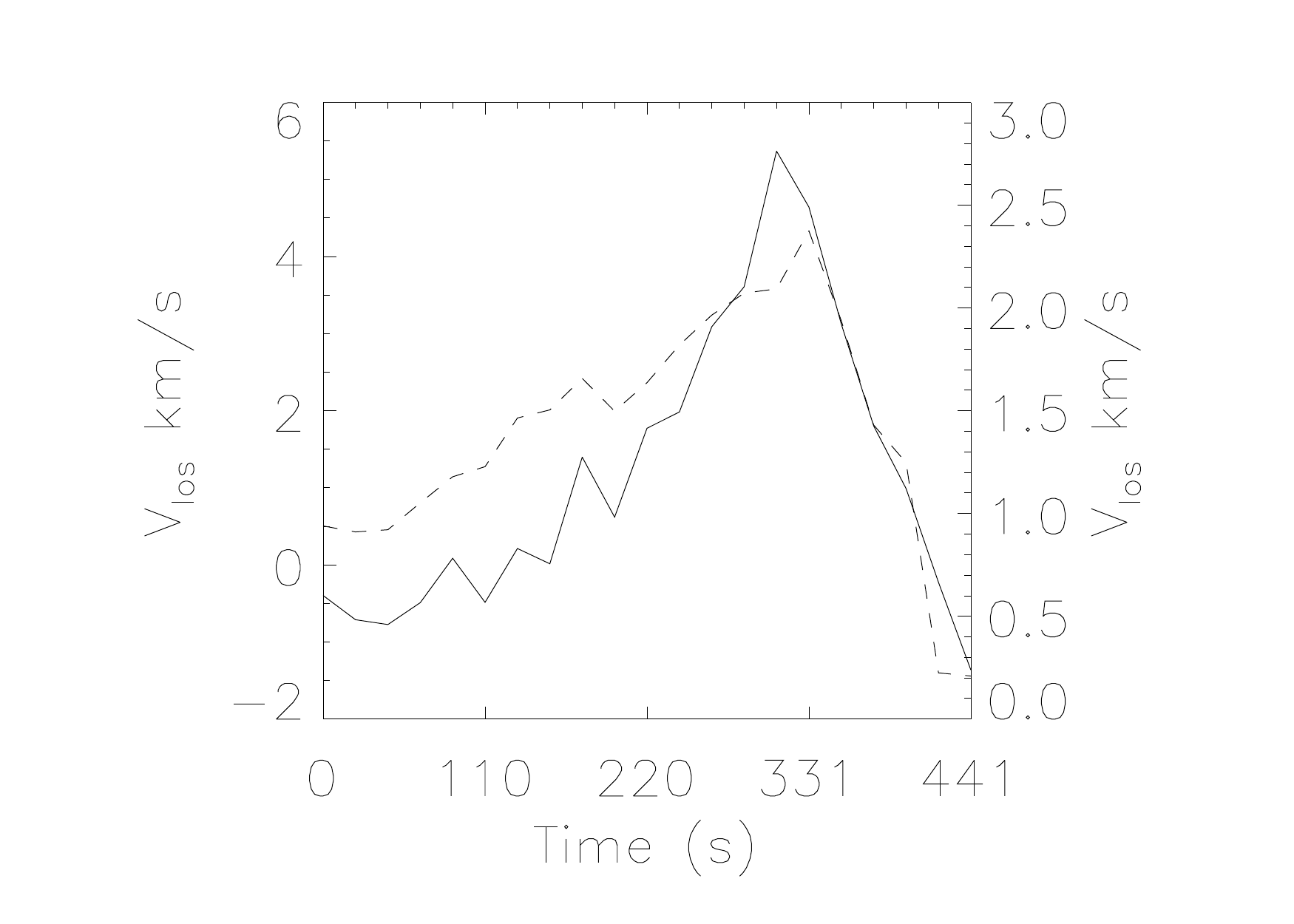}}
\subfigure[Case-c]{
    \includegraphics[bb=55 14 450 323, clip,height=0.18\linewidth]{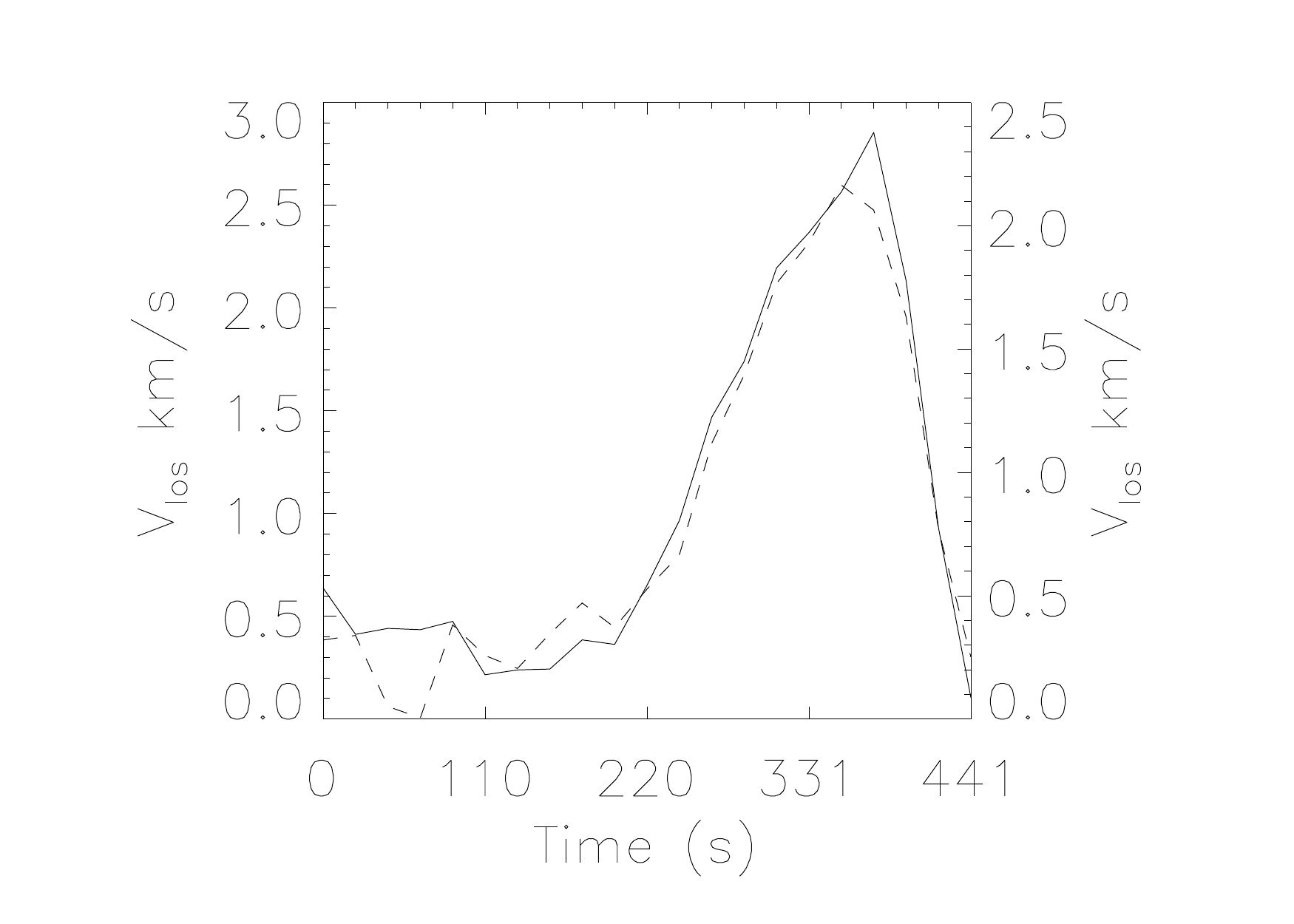}}
 \subfigure[Case-e]{
    \includegraphics[bb=55 14 450 323, clip,height=0.18\linewidth]{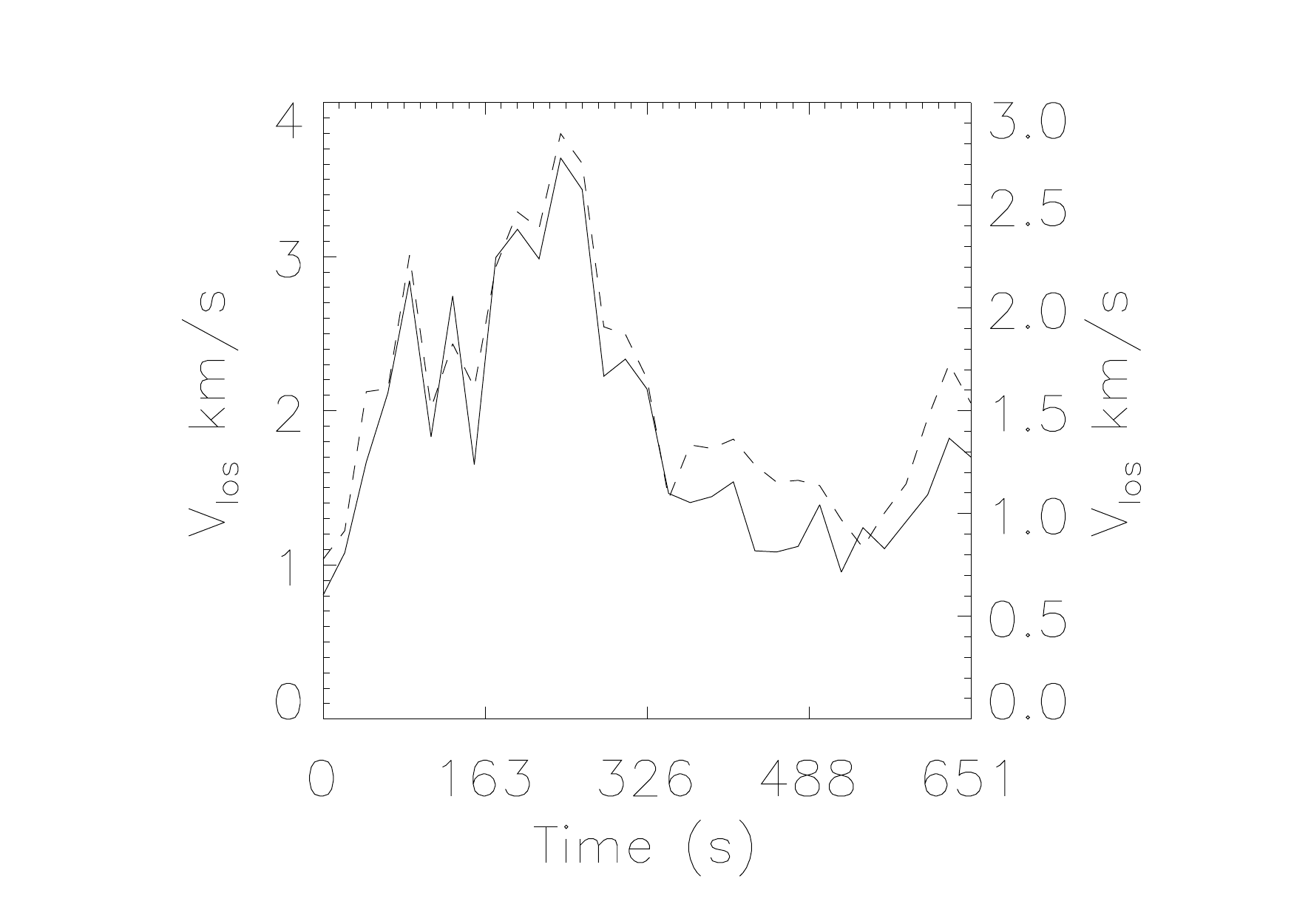}}
    \caption{Time evolution of \vlos\ estimated as the Stokes $V$ zero-crossing velocity (dashed) and obtained from inversions (solid) without stray-light correction for some of the cases. The vertical scale on the left axis corresponds to the Stokes $V$ zero-crossing velocity and the scale on the right axis corresponds to \vlos\ obtained from inversions.}
    \label{}
\end{figure*}
\renewcommand{\thesubfigure}{{}}
\begin{figure*}[htb!]
  \subfigure[ Case-a]{
   \includegraphics[bb=45 95 1133 640,clip,width=0.24\linewidth]{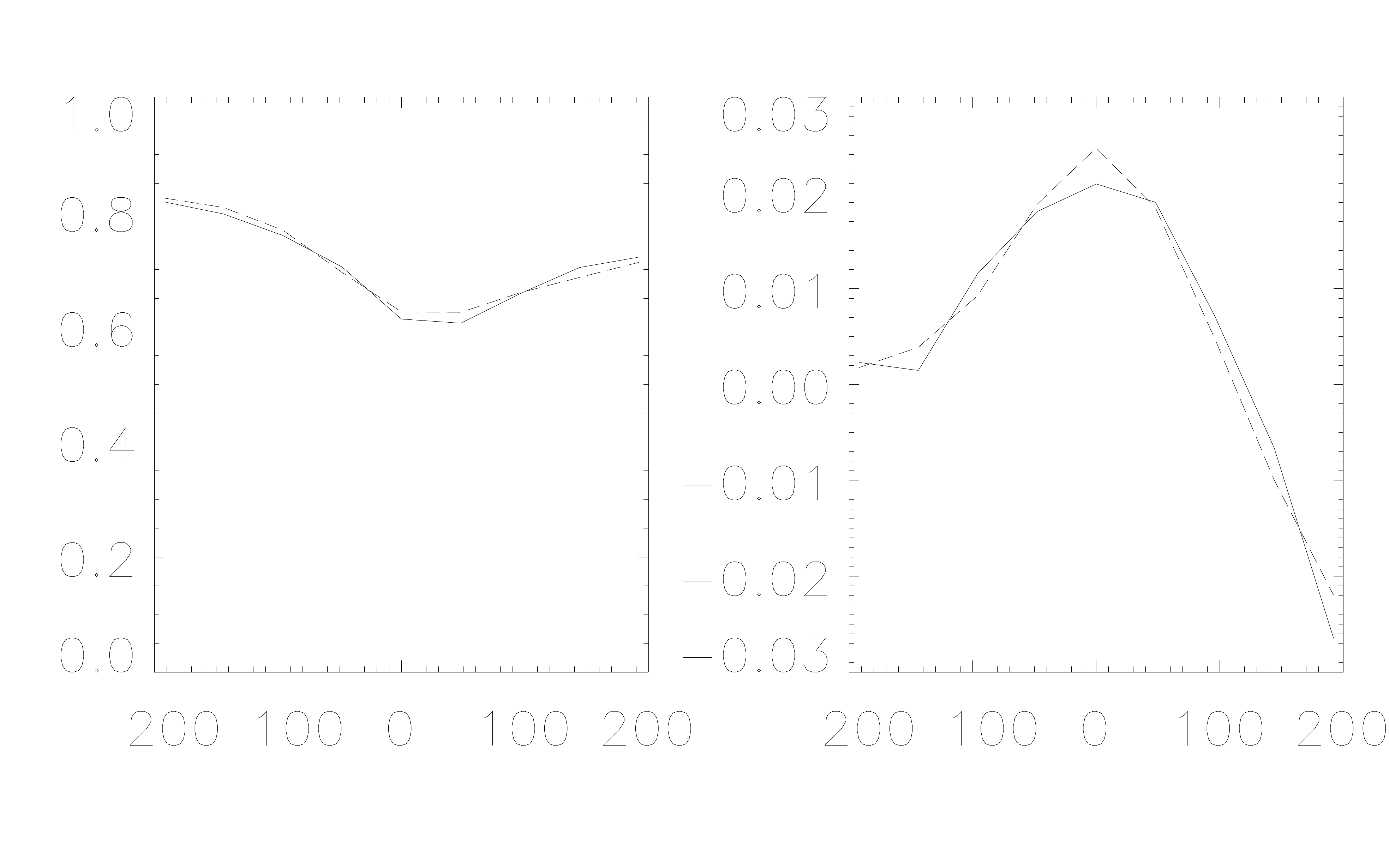}}
  \subfigure[ Case-b]{
   \includegraphics[bb=45 95 1133 640,clip,width=0.24\linewidth]{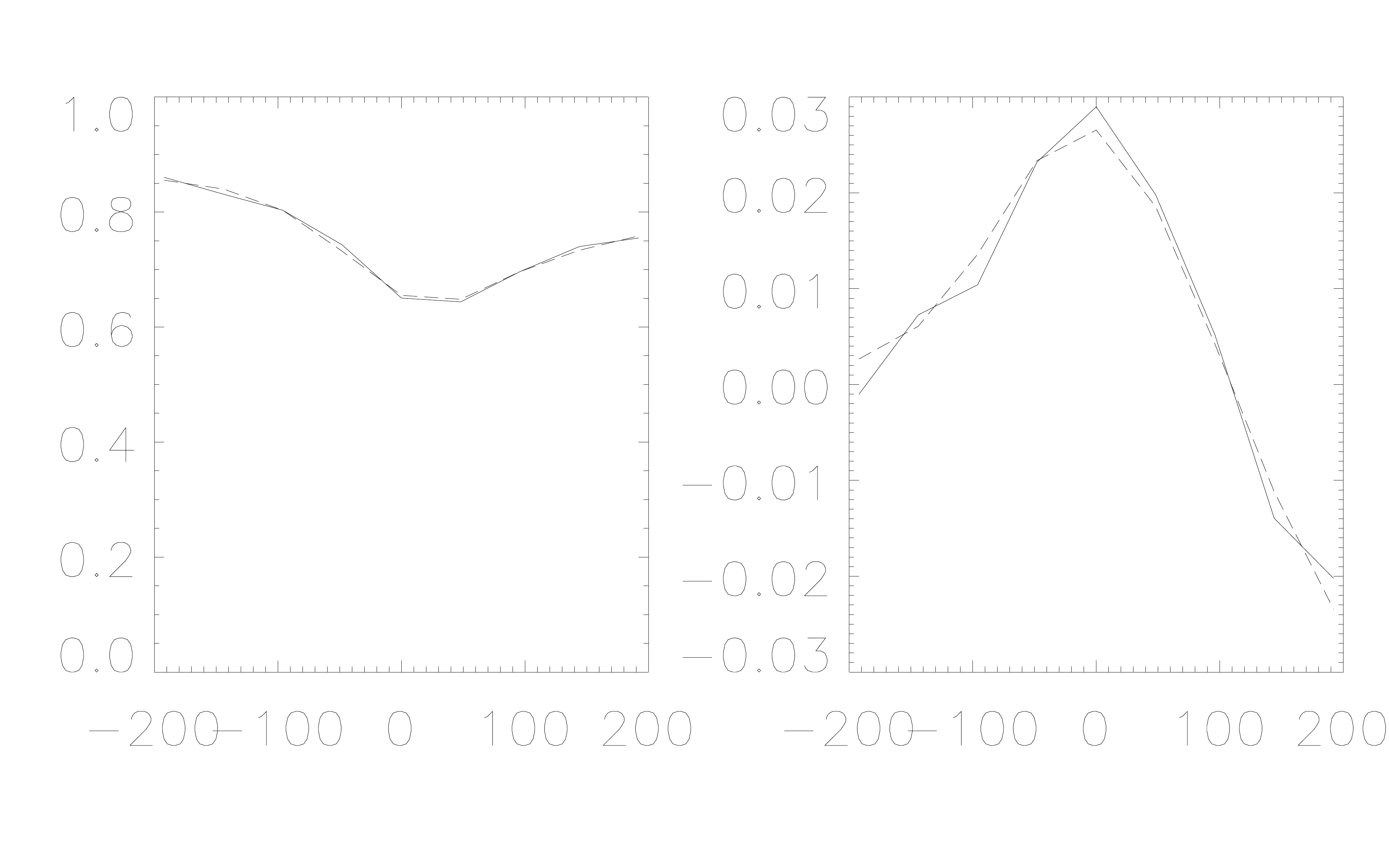}}
  \subfigure[ Case-c]{
   \includegraphics[bb=45 95 1133 640,clip,width=0.24\linewidth]{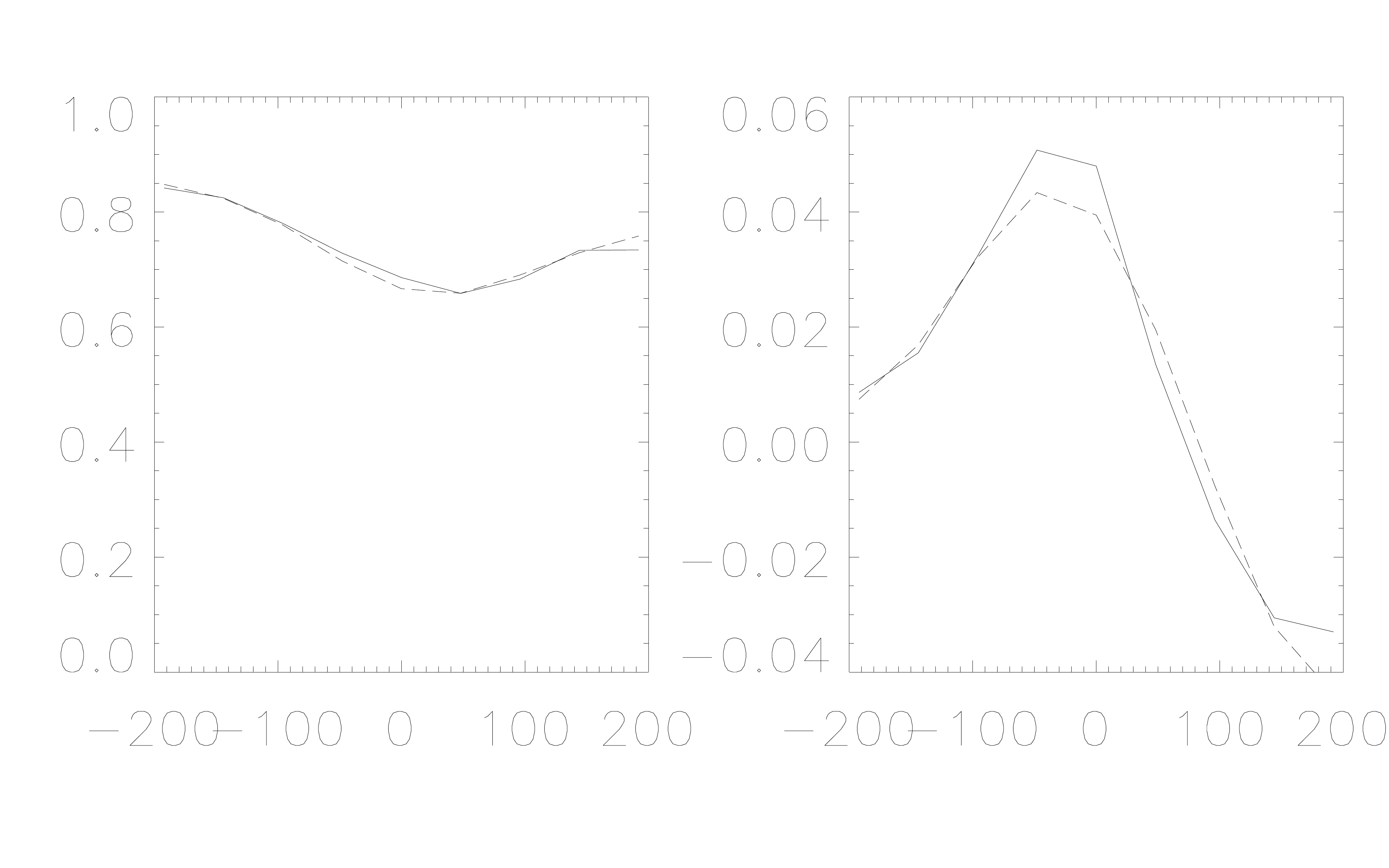}}
   \subfigure[ Case-d]{
   \includegraphics[bb=45 95 1133 640,clip,width=0.24\linewidth]{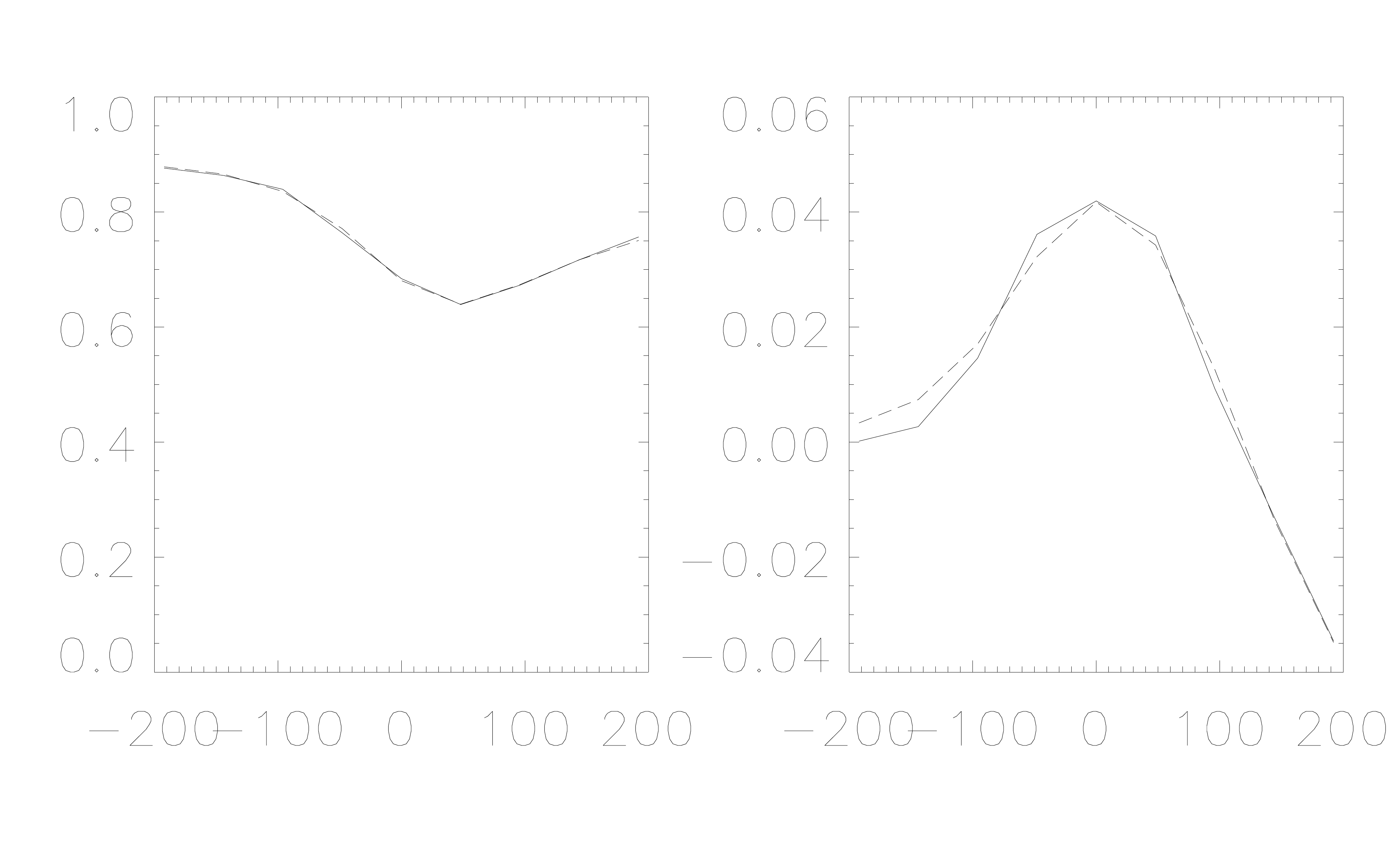}}
   \subfigure[ Case-e]{
   \includegraphics[bb=45 95 1133 640,clip,width=0.24\linewidth]{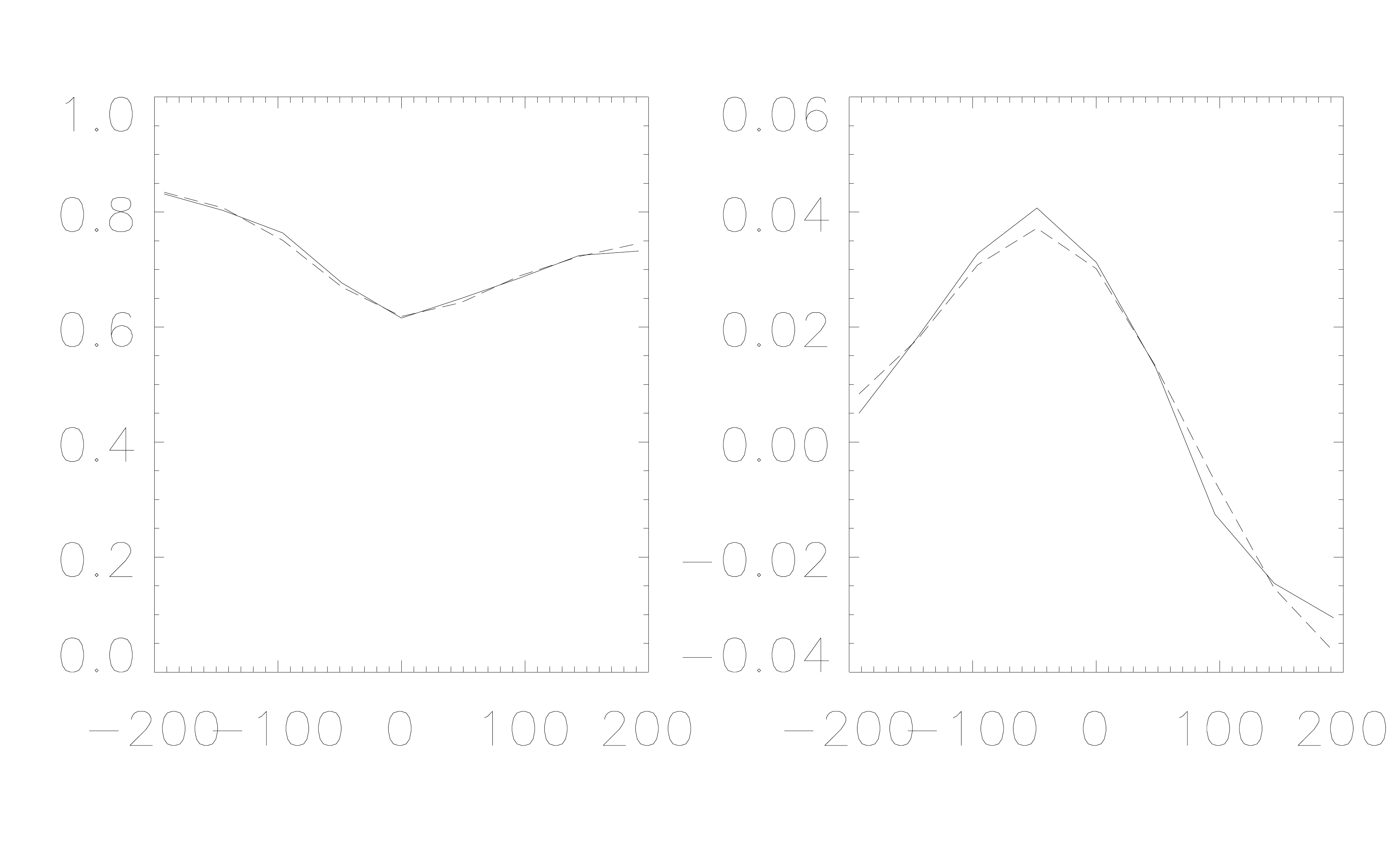}}
   \subfigure[ Case-f]{
   \includegraphics[bb=45 95 1133 640,clip,width=0.24\linewidth]{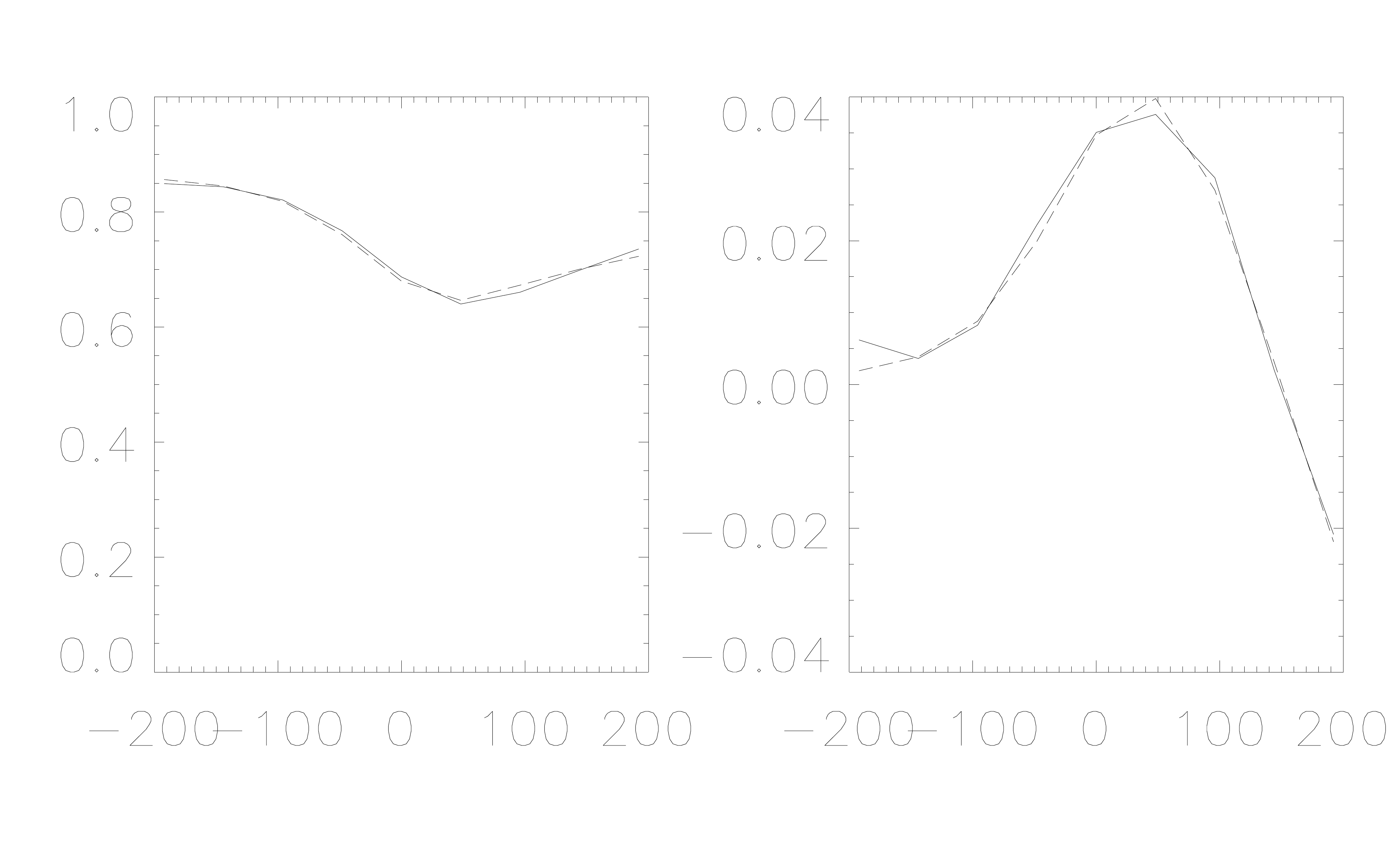}}
    \subfigure[ Case-g]{
   \includegraphics[bb=45 95 1133 640,clip,width=0.24\linewidth]{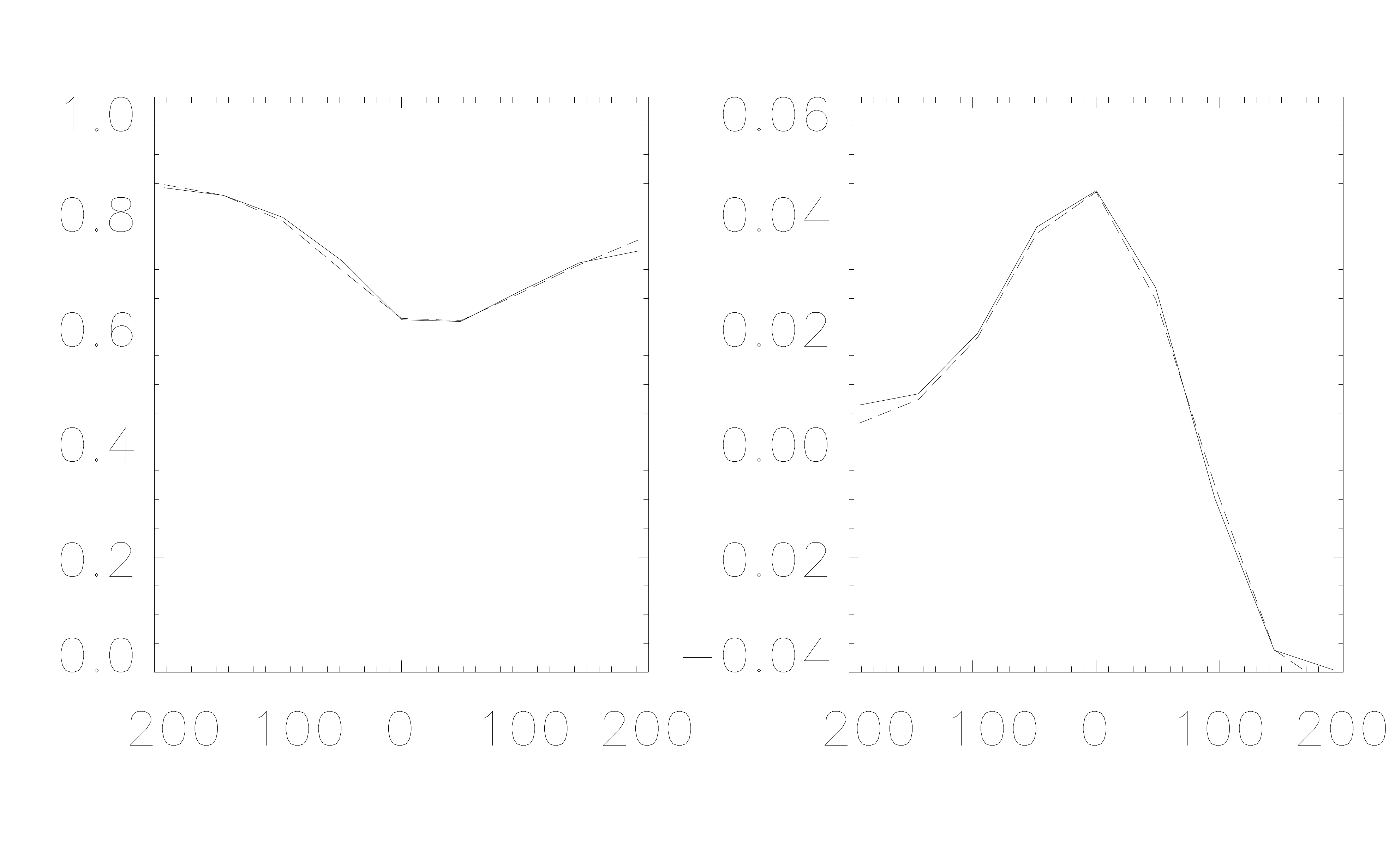}}
  \subfigure[ Case-h]{
    \includegraphics[bb=45 95 1133 640,clip,width=0.24\linewidth]{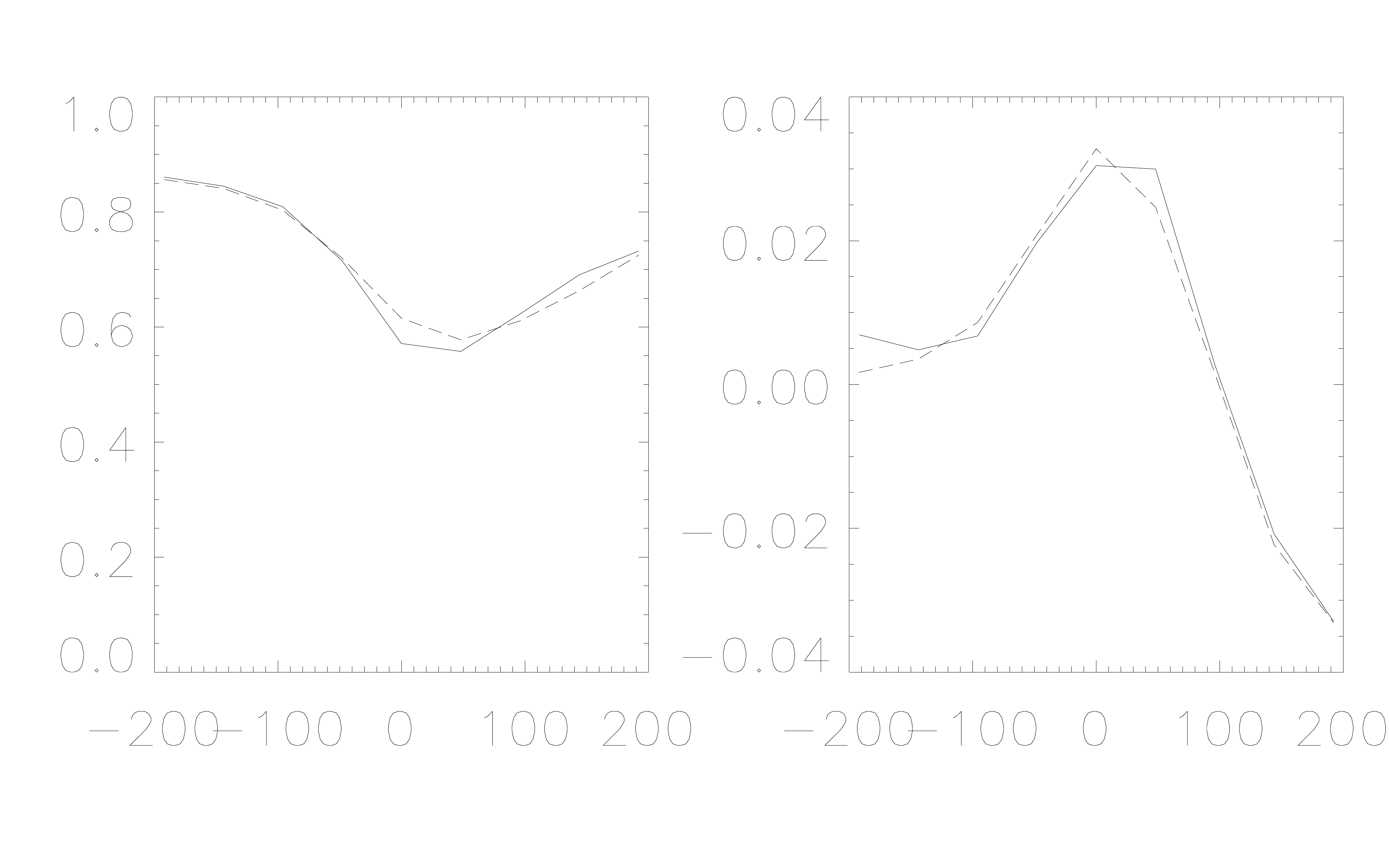}}
   \caption{Observed (solid) and fitted (dashed) Stokes $I$ and $V$
    profiles corresponding to the peak \vlos\ in the different cases with $40\%$ stray-light. Both Stokes $I$ and $V$ are normalized to the continuum intensity, the wavelength of which is outside the range plotted. Wavelengths along the x-axis are in units of m{\AA}.}
  \label{fig:fitobs}
\end{figure*}
\begin{figure*}[htb!]
   \includegraphics[bb=0 0 400 640,angle=90,width=\figwidth]{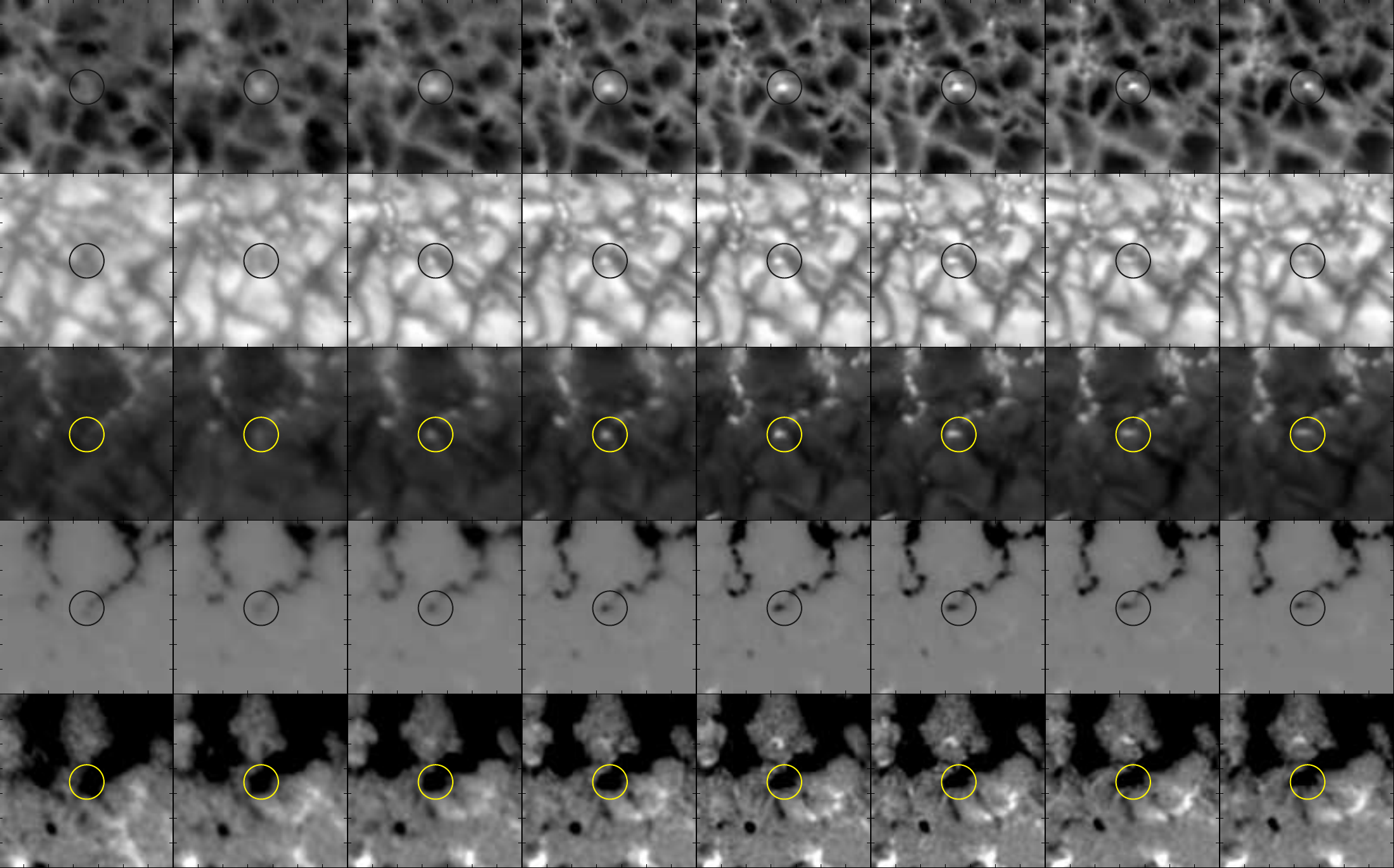}
   \includegraphics[bb=0 0 87 634,width=0.077\textwidth]{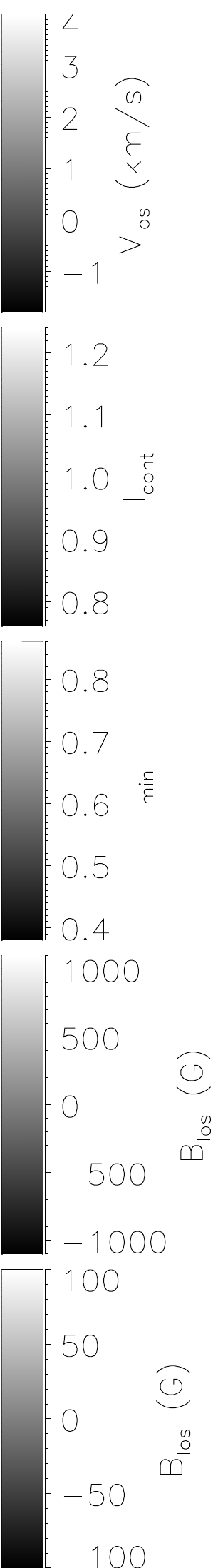}
 \caption{Case $\bf{a}$. Evolution mosaic map. From top to bottom: \vlos\ scaled from $-1.8$~\kmps to $+4.0$~\kmps, $\Ic$ (normalized to the mean continuum intensity), $\Imin$ (normalized to the mean continuum intensity), $\Blos$ scaled from $-1100$~G to $+1100$~G, and $\Blos$ scaled from $-100$~G to $+100$~G. The FOV is 7\farcs17$\times$7\farcs17. Tick-marks are spaced at intervals of $\sim$ 1\arcsec. The snapshots shown (left to right) are recorded at $t=0$, 168, 252, 273, 294, 315, 357 and 378 seconds corresponding to the time in the plots shown in Fig. 7.}
    \label{}
  
\end{figure*}

\section{Analysis}
\label{sec:analysis}
In order to analyze the data, we used a time series of the LOS velocity (\vlos) maps from the inversions. Using the results of the Helix inversions, we identified all downflows in excess of $2$~\kmps observed at the same location in at least two consecutive frames. We found close to 60 events. We selected eight cases from the blocks which had the least seeing variations and where \vlos\ reached $2.3$~\kmps or more. Figure 2 shows observed and fitted Stokes I and V profiles at the time of the peak velocity for the eight cases discussed below. It demonstrates that the fitted Stokes $I$ profiles are slightly red-shifted relative to the observed Stokes $I$ profiles, and the fitted Stokes $V$ profiles are slightly blue-shifted relative to the observed Stokes $V$ profiles. Thus the \vlos\ obtained from the inversions is a compromise between Stokes $I$ and Stokes $V$ velocities {\it explaining the rather poorly fitted Stokes I and V profiles shown in Fig. 2.}

The derived velocities (and magnetic fields) for these small-scale structures are susceptible to stray-light effects. This influence of stray-light is expected to be larger for Stokes $I$ than for Stokes $V$ even if the point spread function (psf) is the same in both cases, since only a small fraction of the field-of-view (FOV) contributes polarized stray-light. To investigate this we estimated velocities from the observed Stokes $V$ zero-crossing shifts by fitting a straight line between the two points closest to zero between the maximum and minimum amplitude positions of the observed Stokes $V$ profiles. At the spectral resolution of CRISP and with the coarse sampling used, this is obviously a crude measure. In Fig. 3 we show the time variation of the LOS velocity obtained from the Helix inversions and the Stokes $V$ zero-crossing velocity for some of the cases. The overall variations obtained with the two methods are similar, but the Stokes $V$ zero-crossing velocity is up to two times larger than that obtained with the Helix inversions.
 
To establish that stray-light indeed explains our poorly fitted Stokes I and V profiles we ran several test inversions with Helix using a fixed stray-light compensation. In this mode a local stray-light profile is computed for every pixel and is modeled as 
\begin{equation}
 I_{obs} = \alpha I_{mag} + ( 1-\alpha ) I_{obs} \ast P
\end{equation}
where $I_{obs}$ is the observed Stokes I, $(1-\alpha)$ is the stray-light fraction, $I_{mag}$ is Stokes I for the (unknown) magnetic component, and $\ast$ denotes convolution with the stray-light psf P, assumed to be a gaussian. We ran several tests with $\alpha$ ranging from 0.3 to 0.99 assuming a FWHM of 40 pixels (2.8 arcsec) for P. The fits with $\alpha=0.99$ are identical to those shown in Fig. 2, showing the relative wavelength shifts between the Stokes I and V profiles, discussed above. Increasing the stray-light (decreasing $\alpha$) leads to an obvious improvement in the fits already with $20\%$ stray-light, but with $40\%$ ($\alpha=0.6$) stray-light the fits look nearly perfect, as shown in Fig. 4.

A choice of $\alpha=0.6$ can be justified by comparing the RMS continuum contrast of the observations with that of 3D simulations. The highest RMS continuum contrast in the present SST/CRISP data is a little over $8\%$ whereas the true contrast (obtained from 3D simulations) at 630 nm is about $14\%$ \citep{2009A&A...503..225W}. This discrepancy is not a question of limited spatial resolution of the SST but quite clearly shows the presence of spatial stray-light \citep{2010A&A...521A..68S}. The precise origin of this stray-light is presently unknown but likely to have contributions from several sources. The measured RMS granulation contrast thus provides a useful constraint on how large the stray-light is. We do not know the detailed shape of the stray-light psf P for the present (or any other) SST/CRISP data, but assuming that it has a width of a few arcsec, its primary effect is to add a spatially slowly varying intensity pattern of low contrast, the effect of which is to reduce the RMS granulation contrast. Based on the above, the CRISP continuum images then suggest a stray-light level of about $40\%$. We therefore re-ran Helix inversions in exactly the same way as described in Sect. 2, but, with a fixed stray-light parameter $\alpha=0.6$.

For both the inversions with and without stray-light, we made evolution mosaic maps using timeseries snapshots of a 7\farcs17$\times$7\farcs17 region surrounding the downflow pixels. Figures 5 and A.1--A.7 show the mosaic maps from the $40\%$ stray-light inversions. We made time-evolution plots of velocities and \Blos\ as shown in Fig. 6 and Fig. 7. To make these plots we first located for each frame the pixel within the marker circle shown in Figs. 5 and A.1--A.7, corresponding to the maximum value of \vlos\ as obtained from the Helix inversions. We then plot the \Blos\ , Stokes V zero crossing and \vlos\ corresponding to this pixel. In the following, we discuss primarily the velocities and magnetic fields obtained with inversions assuming $40\%$ stray-light. We also compare to results obtained without stray-light, in order to demonstrate that the main results obtained are robust.

\begin{figure*}[!htb]
    \subfigure[Case-a]{
    \includegraphics[bb=55 14 450 323, clip,height=0.18\linewidth]{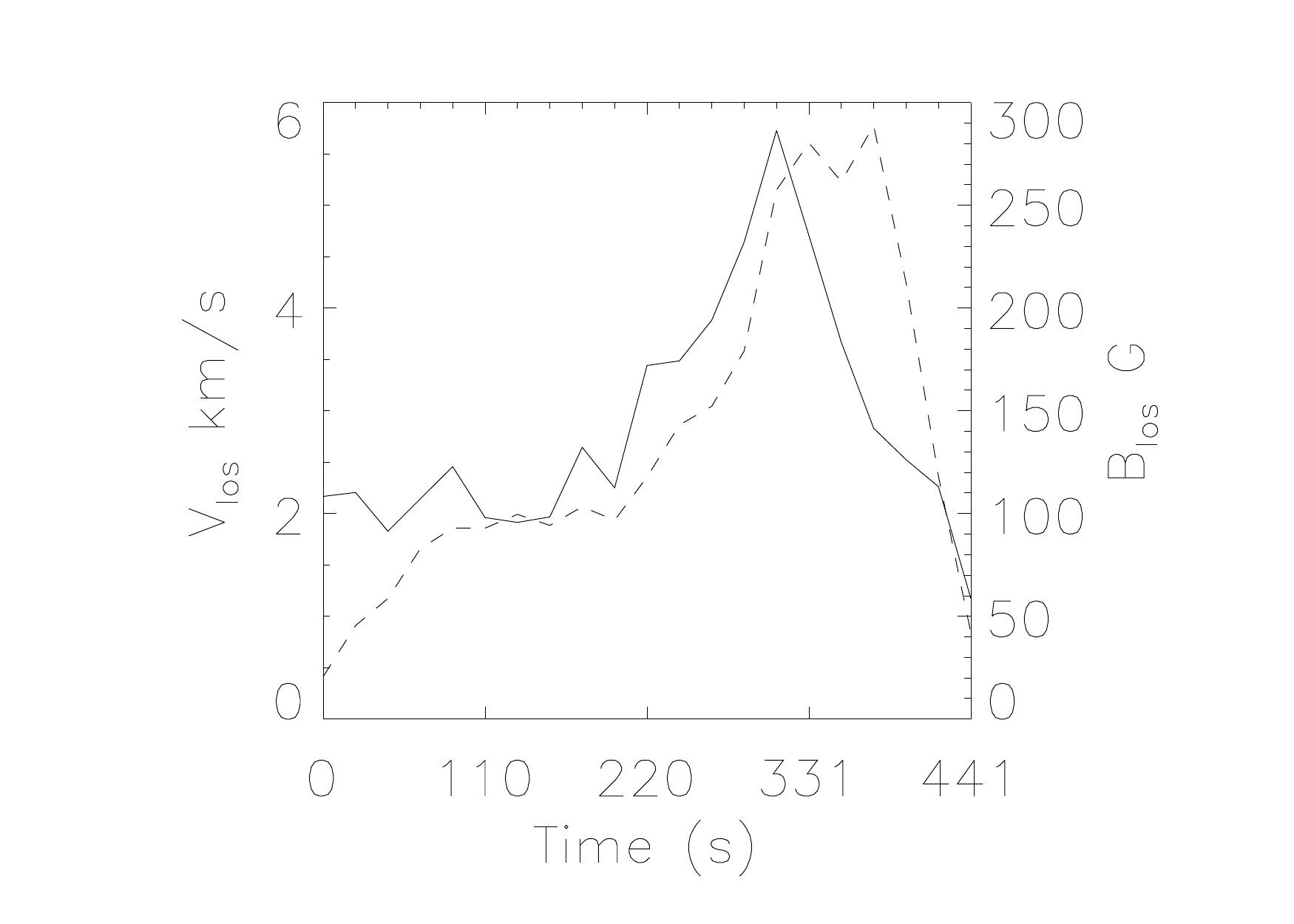}}
\subfigure[Case-b]{
    \includegraphics[bb=55 14 450 323, clip,height=0.18\linewidth]{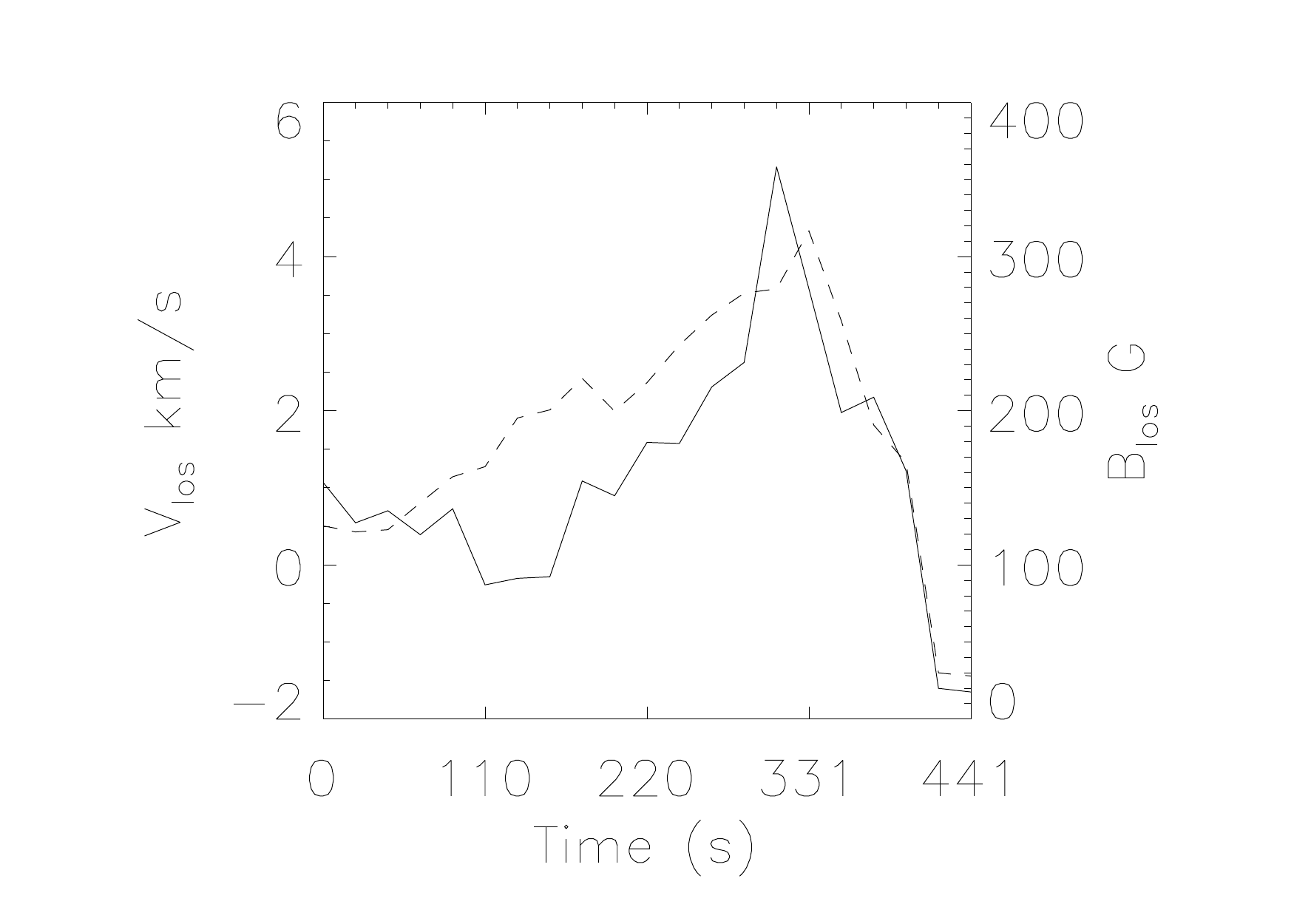}}
\subfigure[Case-c]{
    \includegraphics[bb=55 14 450 323, clip,height=0.18\linewidth]{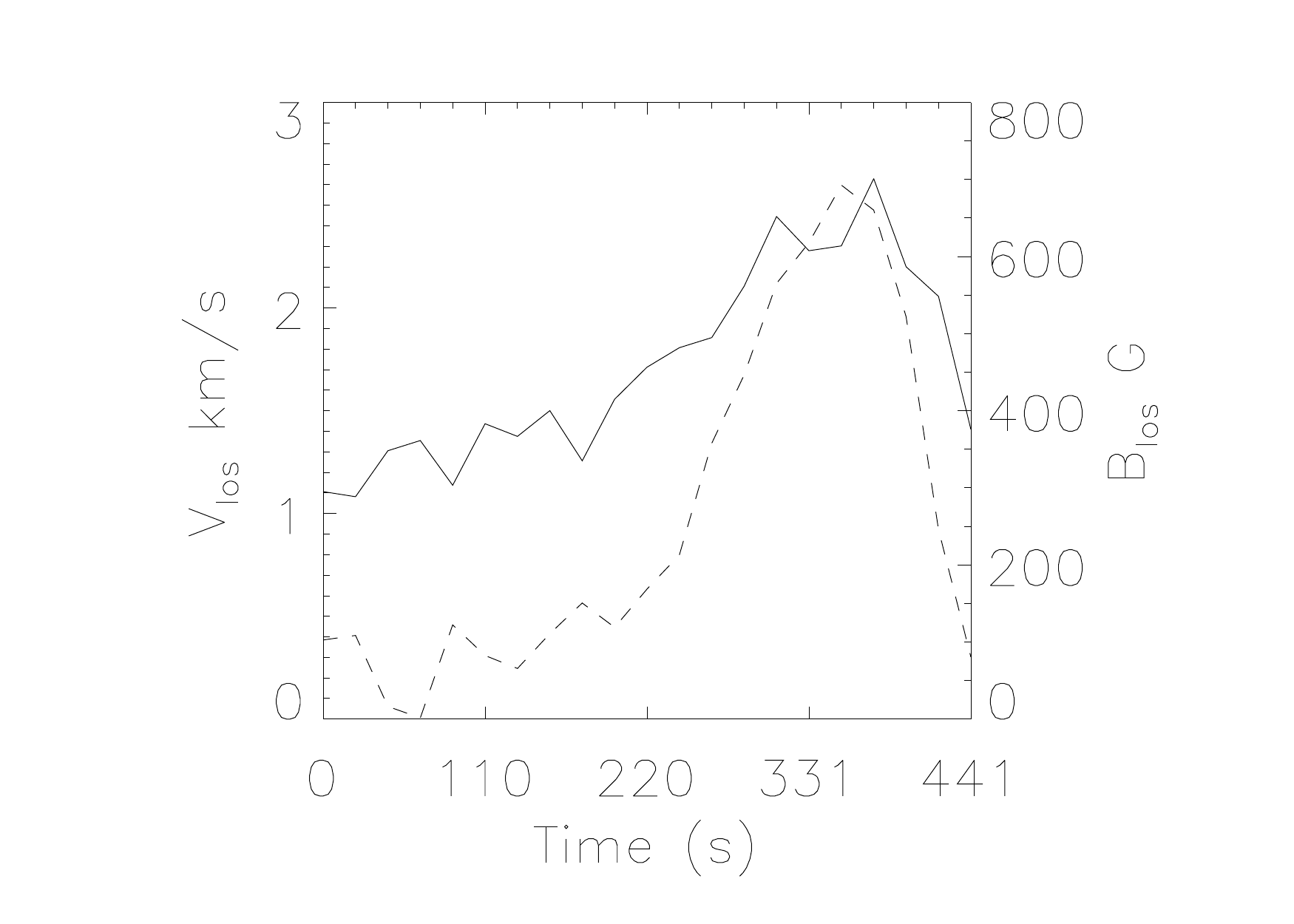}}
\subfigure[Case-d]{
    \includegraphics[bb=55 14 450 323, clip,height=0.18\linewidth]{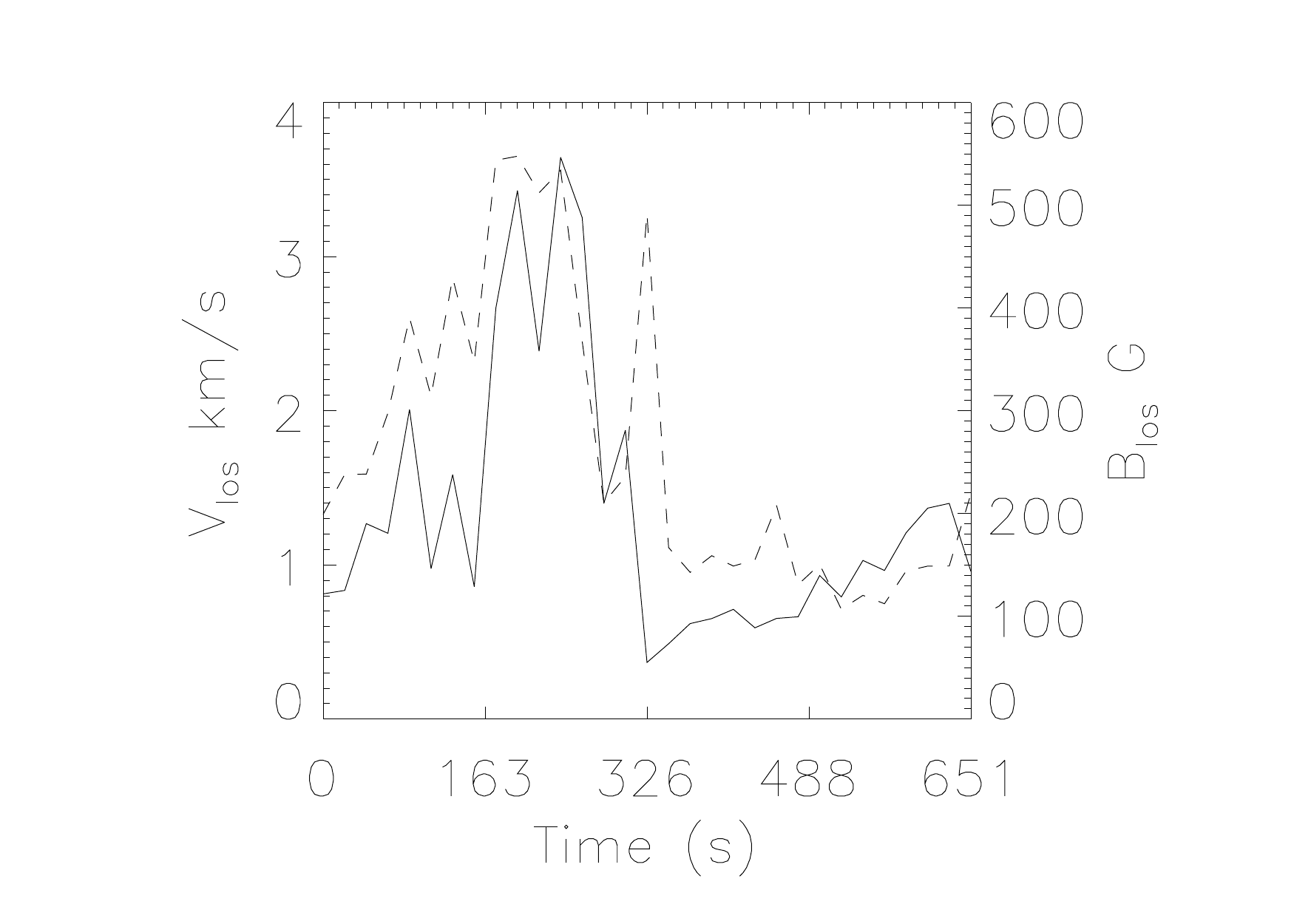}}
\subfigure[Case-e]{
    \includegraphics[bb=55 14 450 323, clip,height=0.18\linewidth]{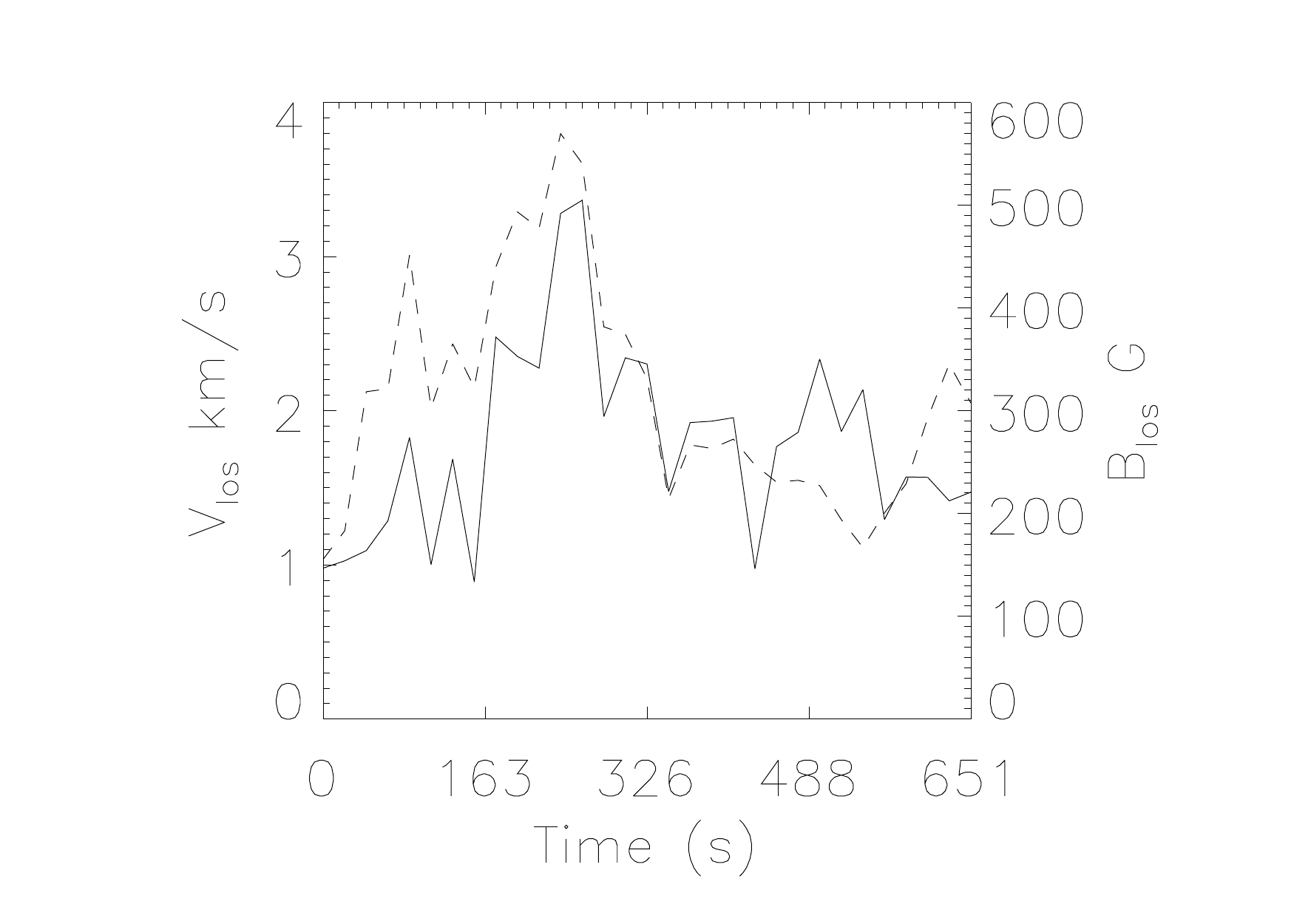}}
\subfigure[Case-f]{
    \includegraphics[bb=30 14 450 323, clip,height=0.18\linewidth]{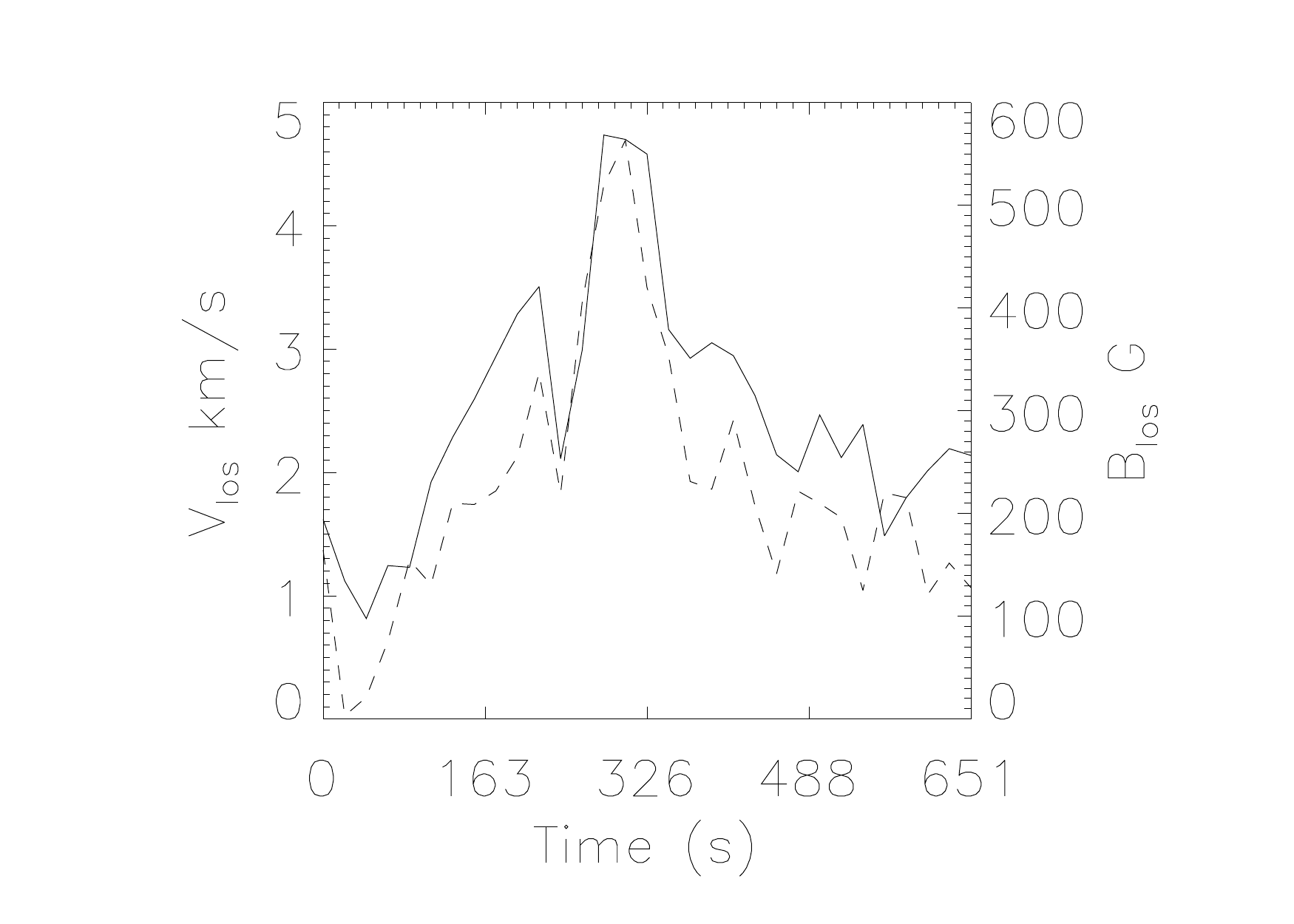}}
\subfigure[Case-g]{
    \includegraphics[bb=30 14 450 323, clip,height=0.18\linewidth]{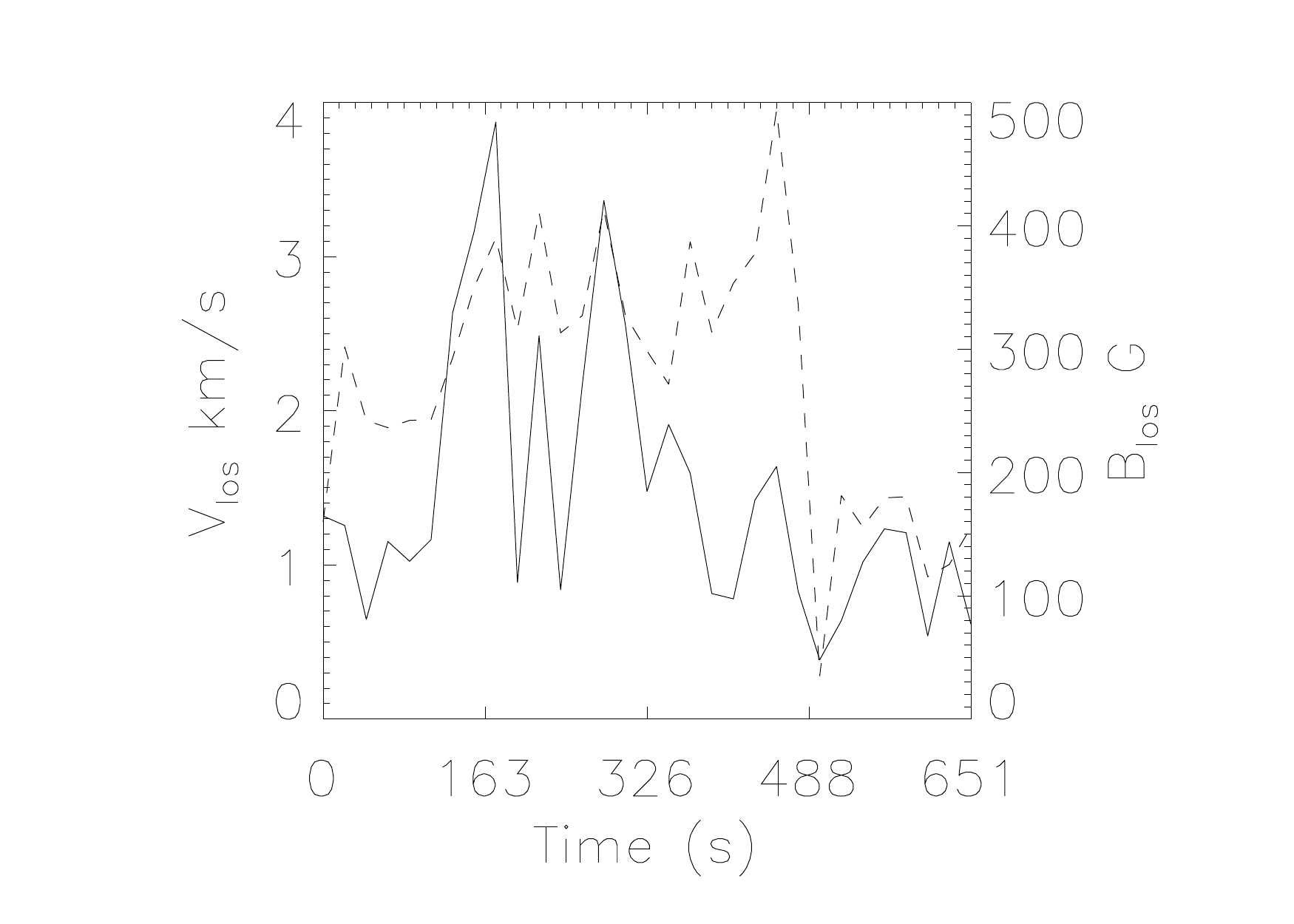}}
\subfigure[Case-h]{
    \includegraphics[bb=30 14 450 323, clip,height=0.18\linewidth]{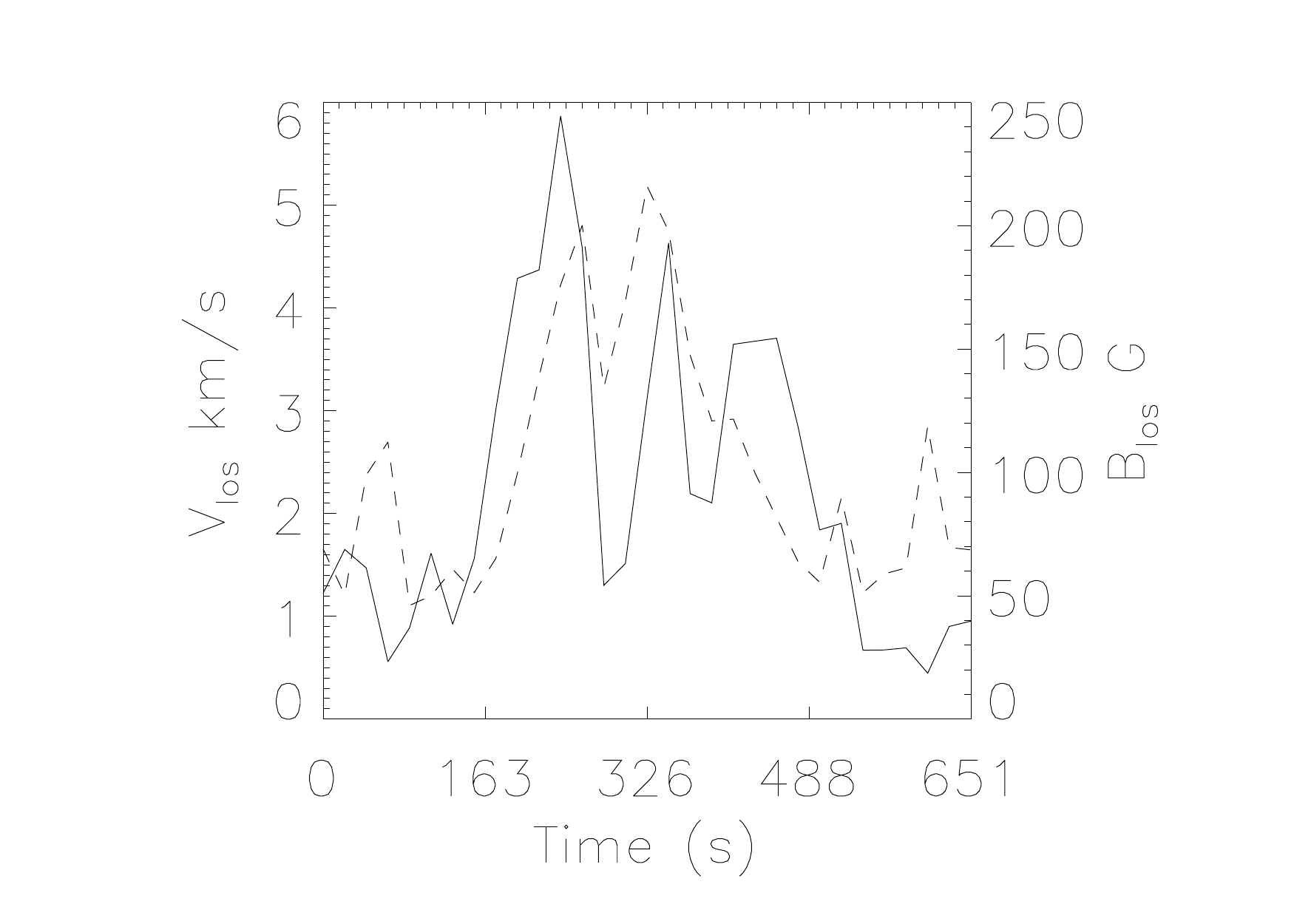}}
    \caption{Time evolution of the Stokes $V$ zero-crossing velocity (dashed) and the strength of the LOS magnetic field $|\Blos|$ (solid) from inversions without stray-light for the different cases. The values plotted correspond to the strongest downflow pixels within the marker circles shown in Figs. 5 and A.1--A.7.}
    \label{}
\end{figure*}

\begin{figure*}[!htb]
    \subfigure[Case-a]{
    \includegraphics[bb=55 14 450 323, clip,height=0.18\linewidth]{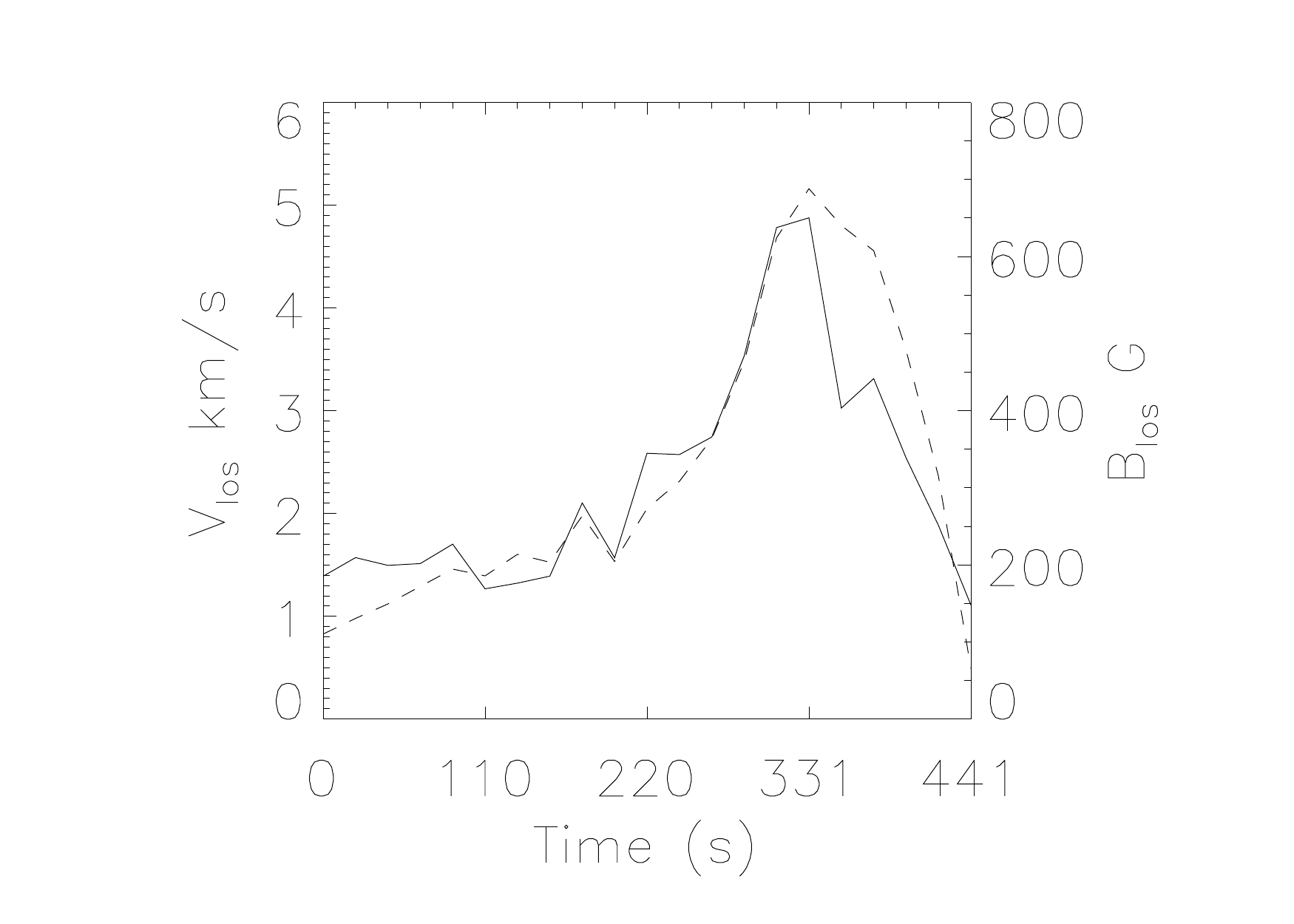}}
\subfigure[Case-b]{
    \includegraphics[bb=55 14 450 323, clip,height=0.18\linewidth]{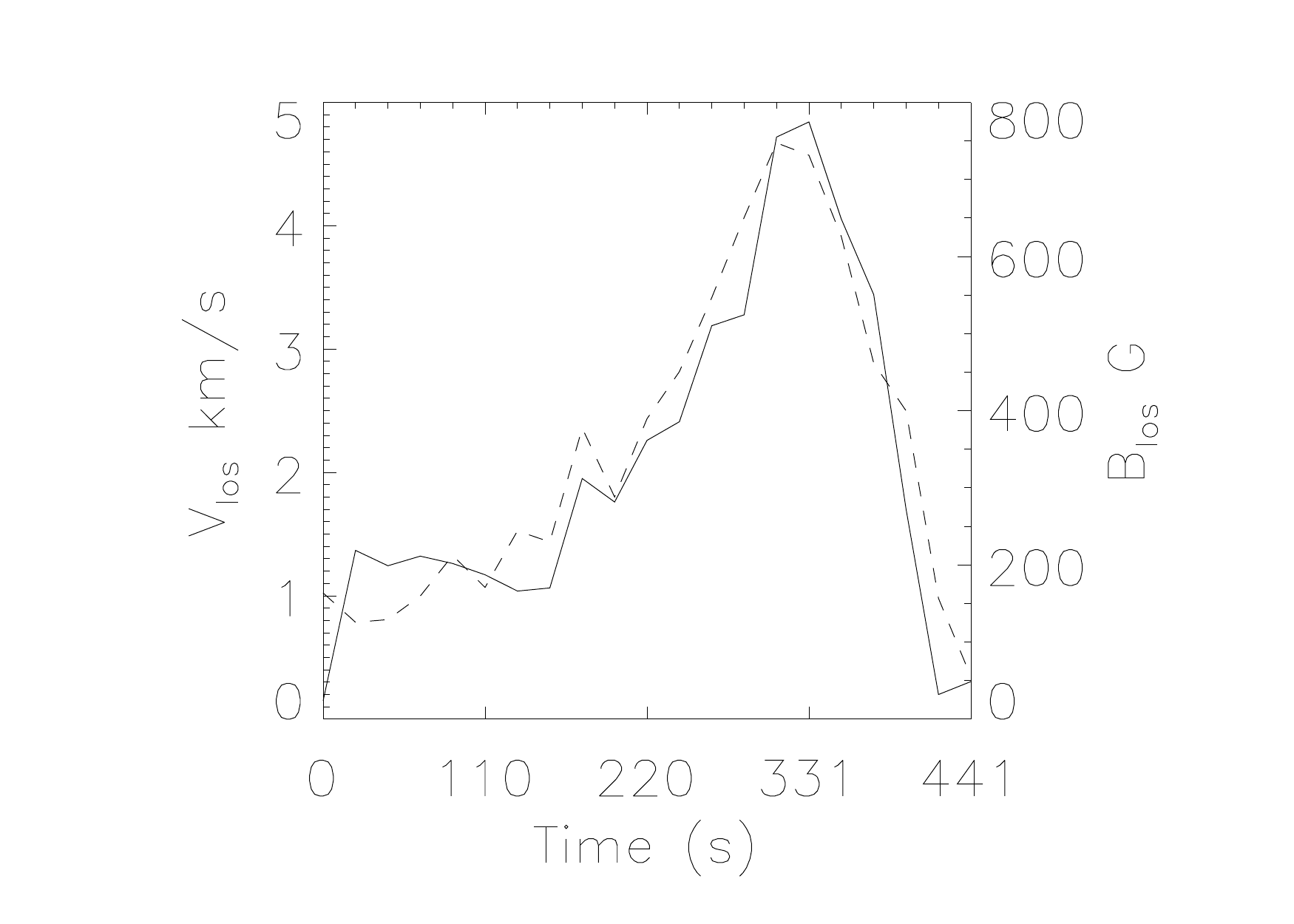}}
\subfigure[Case-c]{
    \includegraphics[bb=55 14 460 323, clip,height=0.18\linewidth]{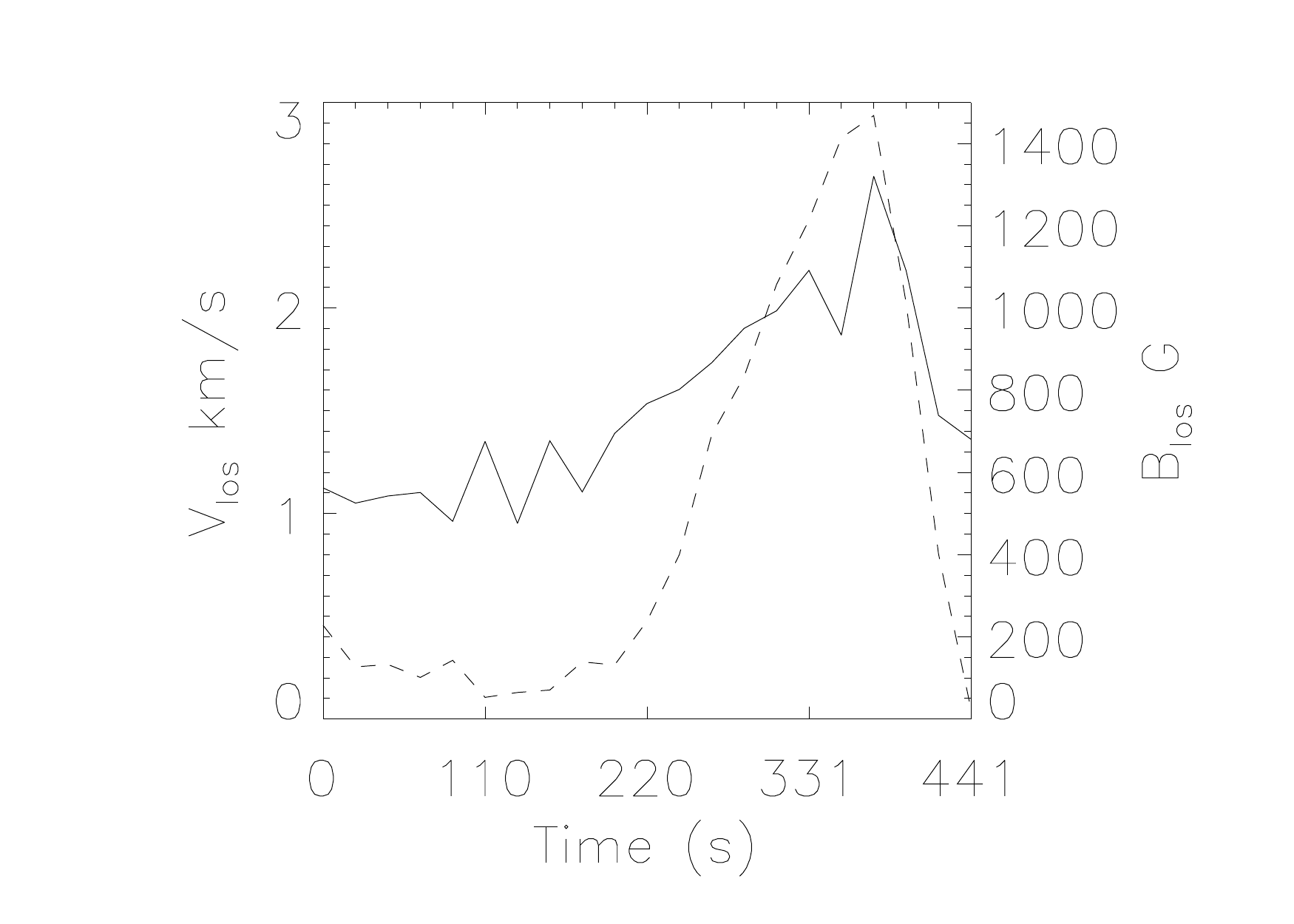}}
\subfigure[Case-d]{
    \includegraphics[bb=55 14 460 323, clip,height=0.18\linewidth]{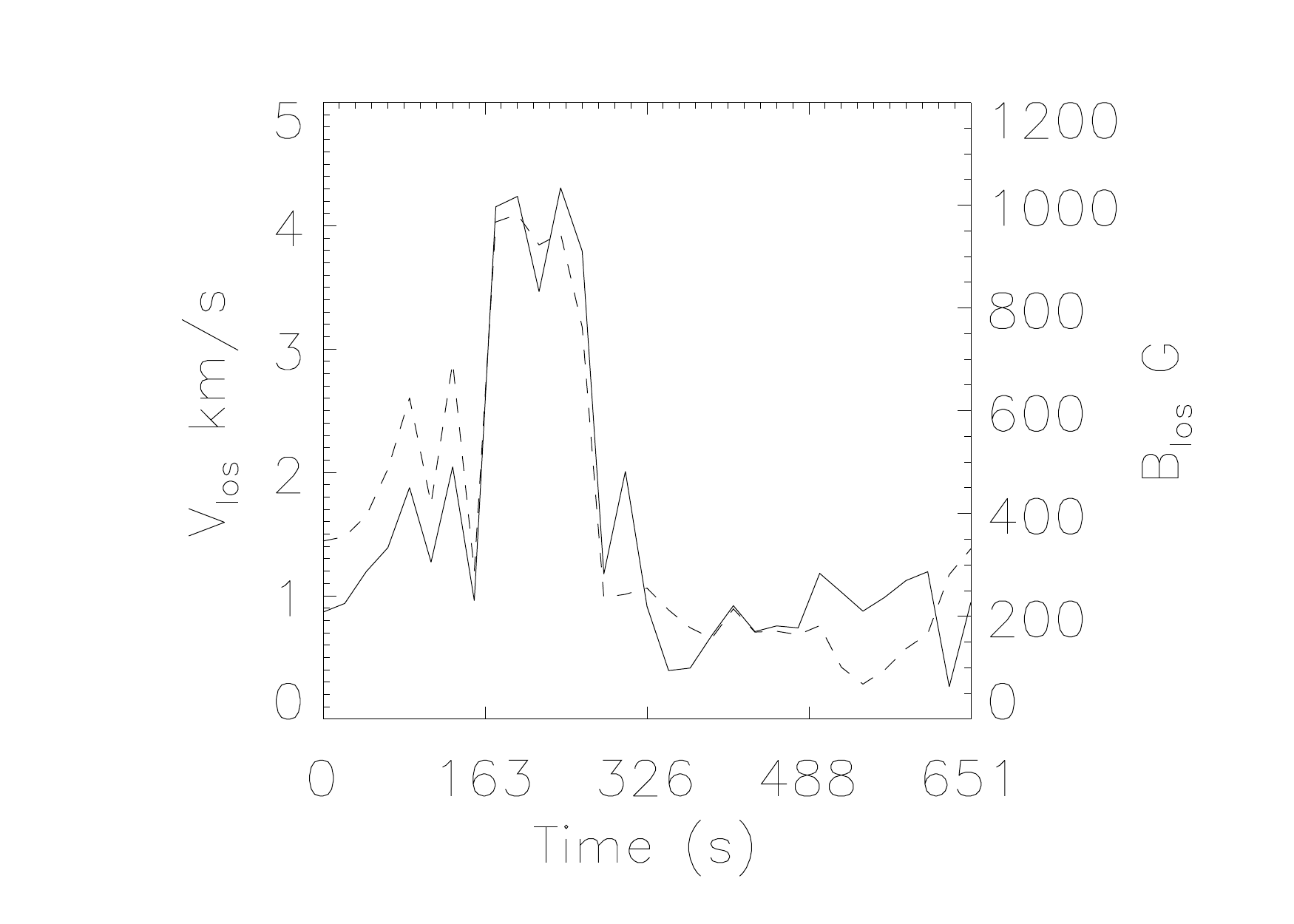}}
\subfigure[Case-e]{
    \includegraphics[bb=55 14 460 323, clip,height=0.18\linewidth]{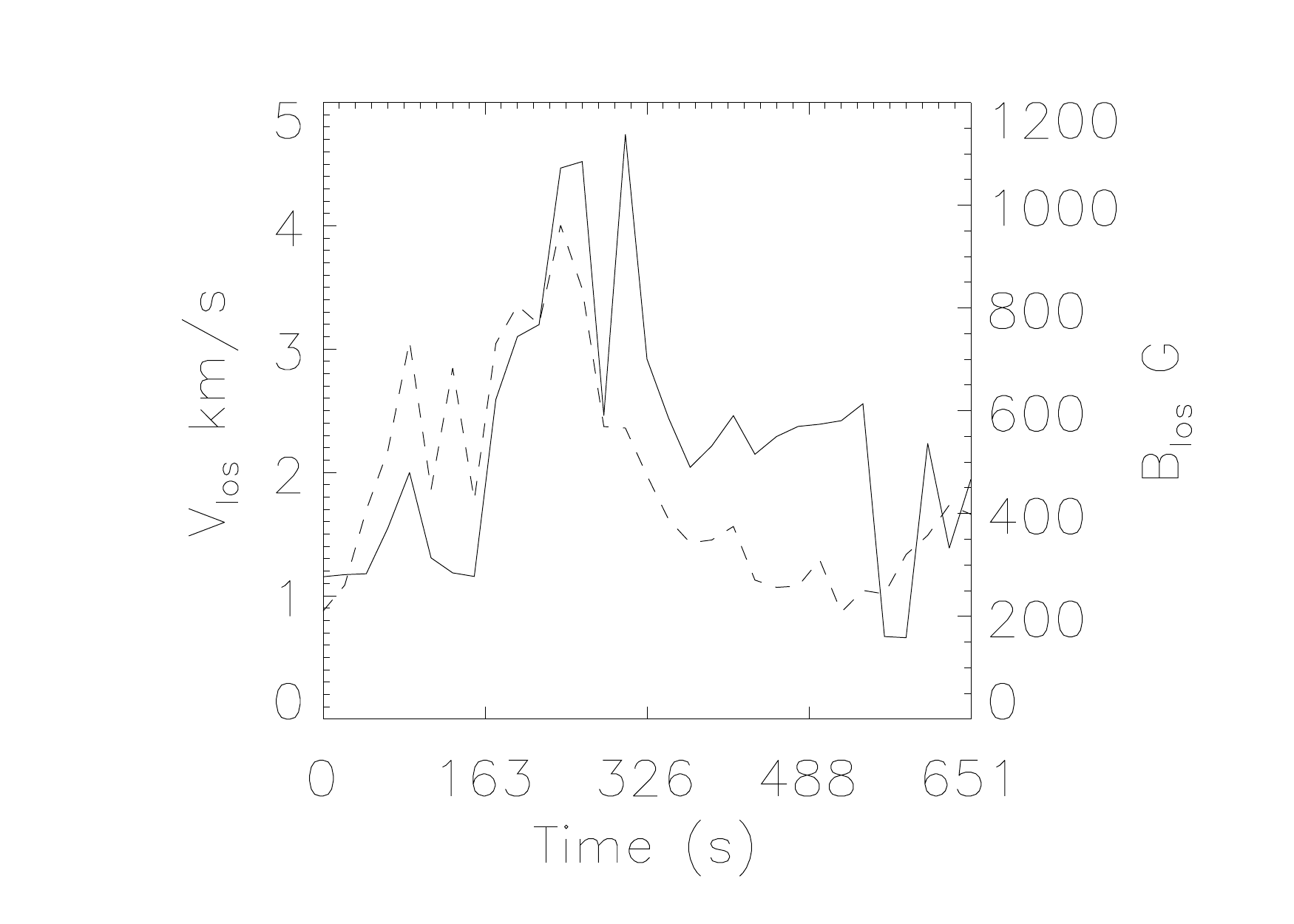}}
\subfigure[Case-f]{
    \includegraphics[bb=55 14 460 323, clip,height=0.18\linewidth]{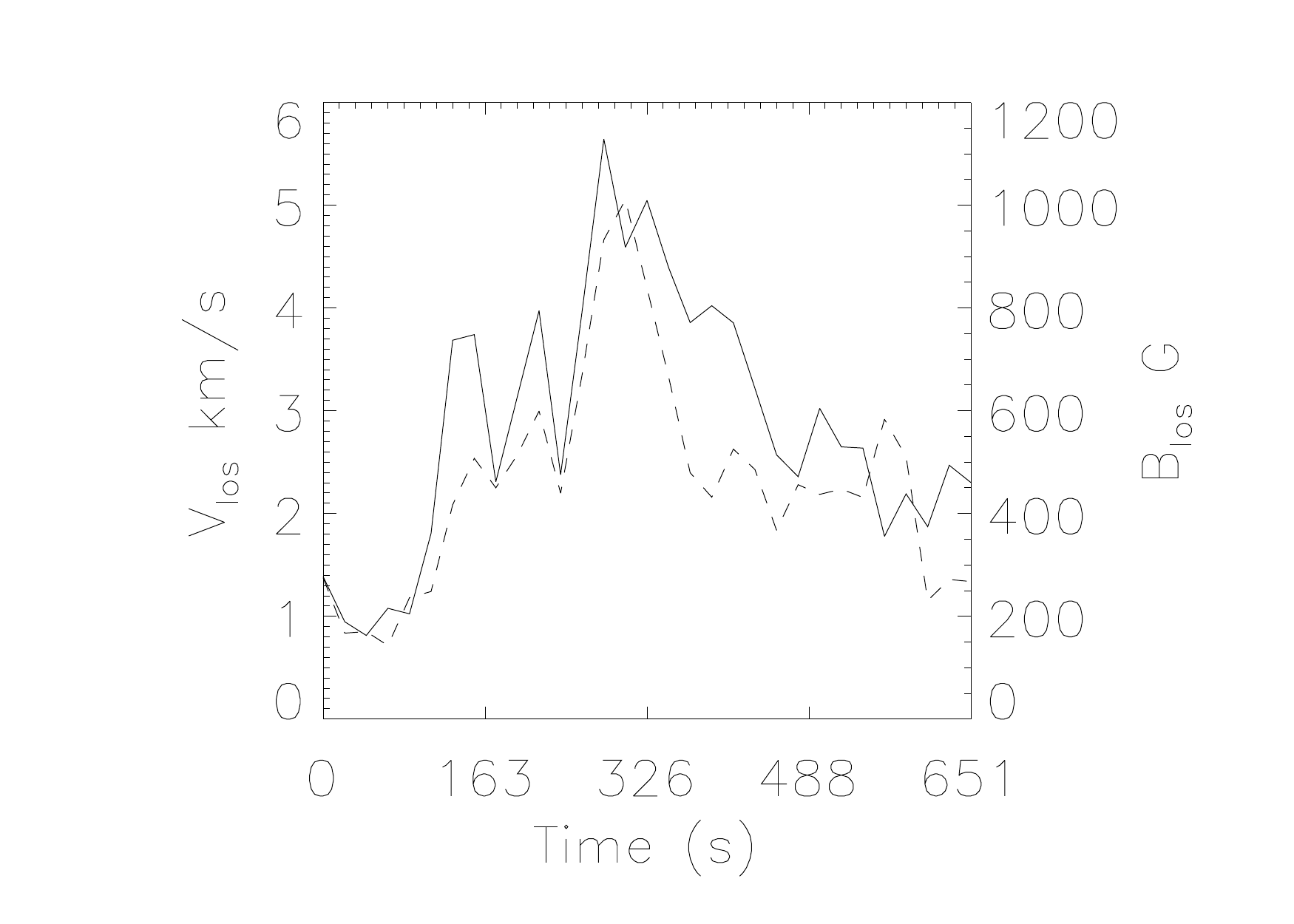}}
\subfigure[Case-g]{
    \includegraphics[bb=30 14 460 323, clip,height=0.18\linewidth]{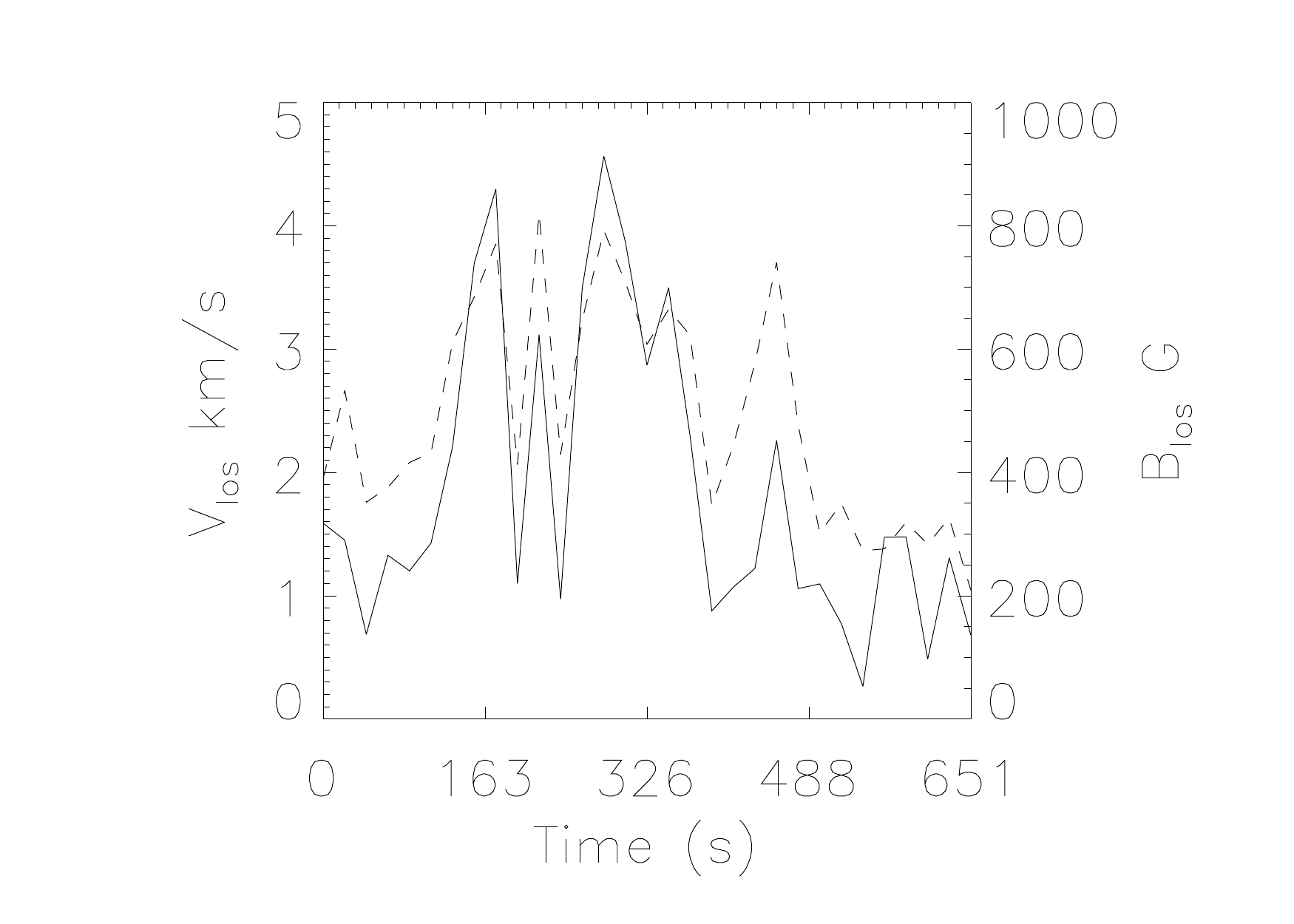}}
\subfigure[Case-h]{
    \includegraphics[bb=30 14 460 323, clip,height=0.18\linewidth]{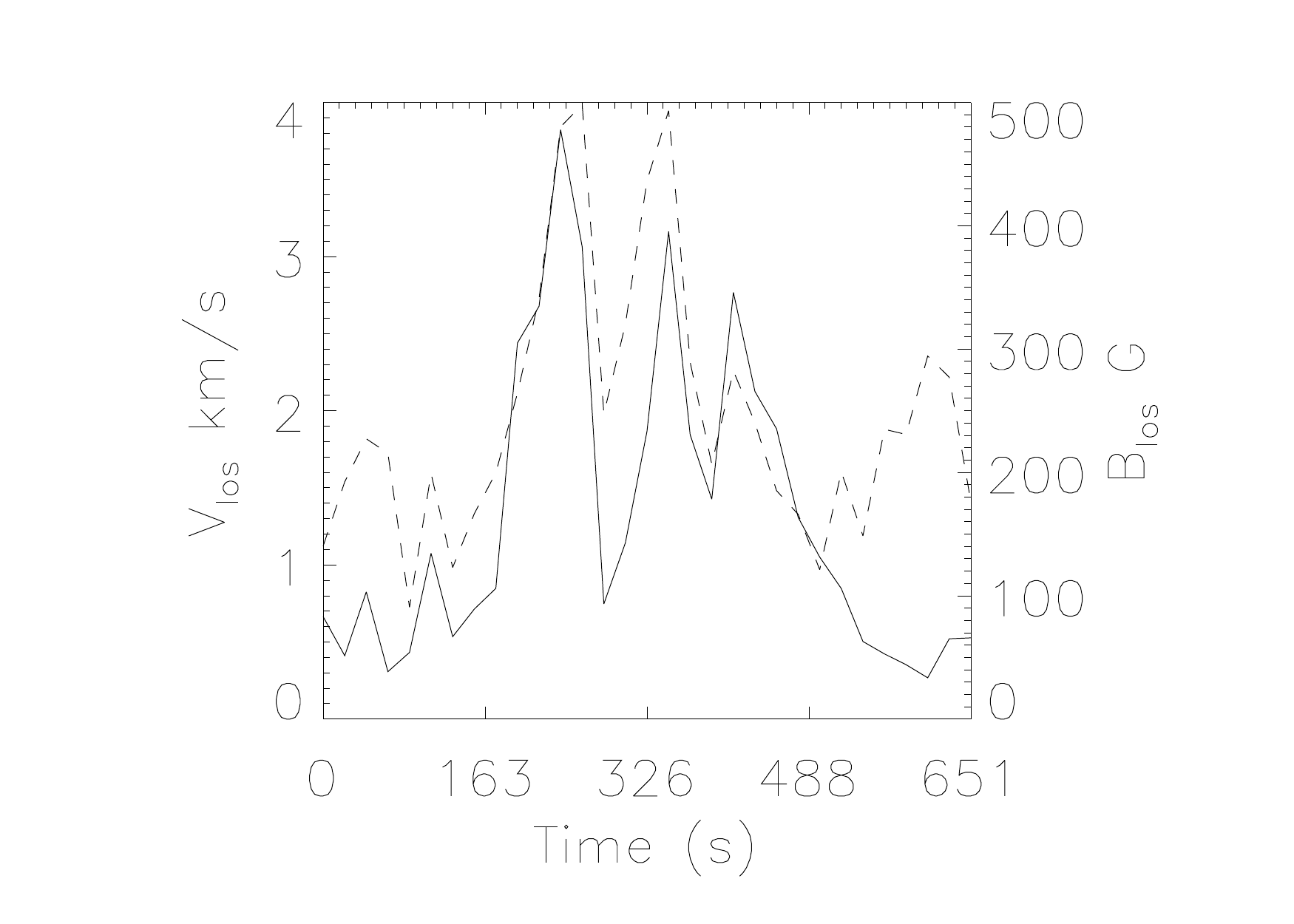}}

    \caption{Time evolution of the \vlos\ (dashed) and the strength of the LOS magnetic field $|\Blos|$ (solid) for the different cases with $40\%$ stray-light. The values plotted correspond to the strongest downflow pixels within the marker circles shown in Figs. 5 and A.1--A.7.}
    \label{}
\end{figure*}

\section{Results}
\begin{table*}[!htb]
\caption{Summary of results for the 8 cases studied.}
    \centering
    \begin{tabular}{  c  c  c  c  c  c  p{8cm} }
    \hline\hline
    Case & \vlos (t=0) & Peak \vlos & $|\Blos|$ (t=0) & Peak $|\Blos|$ & Duration  & Summary \\
      & (\kmps) &  (\kmps) & (gauss) & (gauss) & (s) &  \\ \hline
    a & 1.0 & 5.2 & 210 & 650 & 220 & Starts with the formation of a bright point in an intergranular lane. The bright point evolves into a bright elongated feature.\\ \hline
    b & 0.8 & 4.7 & 220 & 775 & 220 & Appears to start with the formation of a single feature in an intergranular region which splits into two bright points.\\ \hline
    c & 0.3 & 3.0 & 525 & 1320 & 220 & Downflow takes place at the edge of a larger flux concentration which resembles a flower feature in the $\Imin$ map. Bright point is formed at the location of the downflow. \\ \hline
    d & 1.4 & 4.1 & 210 & 1035 & 180 & Starts with a formation of bright feature at a boundary of a large granule. The feature then elongates and transforms into a string feature. The bright point associated with the downflow is formed at the center of the string. The feature then appears to fragment and is pushed out of the field of view due to granular evolution.\\ \hline
    e & 0.9 & 4.0 & 280 & 1140 & 180 & Starts with formation of a bright feature in an intergranular lane which soon elongates and coalesces with nearby bright features to form a single elongated bright feature.  \\ \hline
    f & 1.4 & 5.1 & 280 & 1130 & 240 & A bright point is formed associated with a downflow which takes place at the junction of two string features in an intergranular region. The merger continues leading to a single bright elongated feature which fragments resulting in a big roundish bright feature and a string.\\ \hline
    g & 1.9 & 4.1 & 320 & 915 & 320 &  A bright point is formed in an intergranular lane. The bright point elongates and eventually fragments and disperses.\\ \hline
    h & 1.1 & 4.0 & 85 & 480 & 320 & A continuum bright point is formed in an intergranular lane. The brightening in the corresponding line minimum maps is modest in comparison to all other cases. The point then disappears possibly moving out of the field of view due to granular motions.\\ 
    \hline
    \end{tabular}
 \tablefoot{\vlos\ corresponds to the LOS velocity and $\Blos$ to the LOS flux density obtained from inversions assuming $40\%$ stray-light.}
\label{}
\end{table*}
Comparing Figs. 3, 6, and 7 we find, as expected, that introducing stray-light leads to LOS velocities that are higher than without stray-light and similar to those obtained from Stokes V zero crossing shifts. Also, the estimated field strengths are increased, for many pixels by a factor of two. In all cases studied, there is a transient increase in $\Blos$ (magnetic flux density), accompanying the downflow and the formation of a bright point in both the continuum intensity ($\Ic$) and line-minimum ($\Imin$) maps that is co-spatial with the downflow and field intensification. The downflows reach peak values between $3.0$~\kmps and $5.2$~\kmps. $\Blos$ reaches peak values from 480 G to 1320 G. The case descriptions are summarized in Table 1.

For cases $\bf{a}$, $\bf{b}$, $\bf{d}$, $\bf{e}$, $\bf{f}$, $\bf{g}$, and $\bf{h}$, the location of the peak $\Blos$ coincides with the location of the peak downflow velocity within one pixel (0\farcs071). Case $\bf{c}$ is an exception. Here, the brightening in the $\Ic$ and $\Imin$ maps coincides with the peak downflow pixel. However, the pixel of the peak field is clearly not co-spatial with the strongest downflow. In this case, the downflow occurs close to the boundary of an extended, strong magnetic feature. It may thus be similar to the bright magnetic downflows discussed by \citet{2010arXiv1007.4673N}. 

The duration of these events is in the range 180 to 320 seconds. We find weak opposite polarity field within our marker circle in cases $\bf{b}$, $\bf{g}$, and $\bf{h}$. In case $\bf{b}$ we find weak opposite polarity field between $t=252$ and $t=294$ s. For case $\bf{g}$ we note the presence of opposite field within our marker circle between $t=0$ and $t=42$ s, $t=126$ and $t=168$ s, and between $t=483$ and $t=546$ s. At $t=21$ s (left most column in Fig. A.6) there is an upflow at the location of the opposite polarity field. Upflows are also present in the vicinity of downflows at $t=357$ and $t=441$ s. Also for case $\bf{h}$, we find a nearby weak opposite polarity field between $t=0$ and $t=105$ s, $t=168$ and $t=252$ s and between $t=315$ and $t=462$ s. In cases $\bf{g}$ and $\bf{h}$, we see some rapid fluctuations in the \vlos\ and $\Blos$.

The data presented in Figs. 5 and A.1--A.7 show obvious effects of the variable seeing. Quite clearly, the time variations plotted in Figs. 6 and 7 are affected by these seeing variations. In Fig. 5 we see several small-scale downflow sites that could be argued to suggest that the velocity variations are only artifacts of seeing. However, the line minimum intensity variations clearly show the formation of a bright point at the downflow site whereas other bright structures show much less changes over time, in spite of the variable seeing. We therefore conclude that the \emph{overall} temporal variations in \vlos\ and $\Blos$ are real, rather than artifacts.

\section{Discussion}

The eight cases of strong, transient downflows discussed in the present paper were identified from a time sequence of polarimetric data using the strength of the downflow and the consistency of the detection during two consecutive frames as the \emph{only}  selection criteria. We find that all downflows thus identified a) are associated with magnetic field, i.e., they do not occur in field-free regions,  b) correspond to a \emph{transient} intensification of the measured magnetic flux density and c) are co-spatial with the formation of a short-lived bright point at continuum and line minimum wavelengths. These signatures are in overall agreement with an interpretation in terms of flux-tube collapse \citep{1978ApJ...221..368P, 1979SoPh...61..363S}. However, we emphasize that none of our observed cases seem to lead to a \emph{permanent} strengthening of the magnetic field. Already \citet{1998A&A...337..928G} presented 2D MHD simulations demonstrating that convective collapse can lead to the formation of both stable and unstable kG flux sheets. They found that when the initial magnetic flux is large, the obtained supersonic downflow turns to strong upflow after rebounding against the dense bottom of the flux tube. The upflow causes the flux tube to disperse. \citet{2001ApJ...560.1010B} reported spectropolarimetric observational evidence for magnetic element evolution and destruction in a similar framework. In our observations, we do not see evidence for such a rebounding upflow except for case $\bf{g}$, but for cases $\bf{a}$, $\bf{b}$, and $\bf{c}$, the time span of our observations may be too short to include such upflows. Another reason for the missing upflows may be that the total magnetic flux of some of our flux concentrations is too small. When initiating the simulations with only a small amount of magnetic flux, \citet{1998A&A...337..928G} found that the convective collapse was less efficient with subsonic downflows reaching $4$~\kmps and no flow reversals taking place. The flux sheet formed was small, and the field strength only 1 kG at $\tau=0.1$. They attributed the lower efficiency of convective collapse of the smaller flux sheet formed to efficient radiative energy exchange with the surroundings \citep{1986Natur.322..156V,2000ApJ...544..522R}. \citet{1998A&A...337..928G} also predicted that in order to observe convective collapse of thick flux tubes, a spatial resolution of 0\farcs5 is sufficient. However, to observe the collapse of \emph{small} flux tubes, 0\farcs1 resolution is needed.

It is quite clear that the high spatial resolution of our data is crucial in detecting some of these events, in particular as regards the bright-point formation. In a recent paper, \citet{2010A&A...509A..76D} discuss 3D MHD simulations of magnetic field intensification and compare the original simulations with simulations degraded to Hinode resolution. They present three cases of convective collapse, out of which two cases involve small-size features. For the larger feature at the original resolution, a supersonic downflow is followed by a strong upflow. Degrading to Hinode resolution, they see a downflow of the order of $5$~\kmps and an upflow of only $0.5$~\kmps. The magnetic field at the original resolution increases from 250 G to about 1.6 kG and then drops to a sub kG value at the end of the event after 400 s. The maximum continuum brightness is reached 100 s after the peaking of the velocity. At the Hinode resolution, the apparent field strength (or, flux density) is lower by as much as a factor of 10. The obtained continuum intensity at degraded spatial resolution shows a poor correlation with that obtained at the original resolution and only the general trend agrees. For the smaller flux concentrations, \citet{2010A&A...509A..76D} observe a simultaneous increase of the velocity, flux density, and continuum brightness. The downflows are weaker in comparison to the previous case described. At the degraded spatial resolution, the downflows "observed" are even smaller ($3$~\kmps and $2$~\kmps),  with no signs of flow reversals. The duration of these events is also shorter with the downflows suppressed after 280 s. In both these cases, the field strength does not reach kG values when the \vlos\ peaks (at the original resolution). The authors conclude, in agreement with  \citet{1998A&A...337..928G},  that lateral radiative heating of the \emph{smaller} features is responsible for the lower magnetic field reached for these cases. They also conclude that Hinode would completely miss the bright-point formation in this case due to inadequate spatial resolution.

\citet{2008ApJ...677L.145N} presented observational evidence of convective collapse with polarimetric Fe I 6302 data from Hinode. They observed the formation of a bright feature associated with a strong downflow and simultaneous field amplification. The peak velocity derived from Stokes $V$ zero-crossing shifts reached close to $6.0$~\kmps and the field strength increased from an initial value of 400 G to 2 kG and then dropping to 1.5 kG. The filling factor measured remained unchanged at 0.5. This observation is different from ours by apparently resulting in a flux tube that remains in a strong-field state. There is a transformation of the downflow to an upflow of about $2.0$~\kmps, which might be explained by a relatively \emph{large} flux concentration being involved in the collapse and also explaining the higher field strengths (flux density multiplied by filling factor) obtained, consistent with the simulations of \citet{2010A&A...509A..76D} and \citet{1998A&A...337..928G}. In Hinode observations of a quiet region away from a sunspot, \citet{2008ApJ...680.1467S} found a strong downflow associated with the formation of a G-band bright point. The downflow reaches supersonic speed, $7.8$~\kmps, but there are no reported flow reversals.  During this event, the field strength increases from 479 G to 1514 G and the authors interpret this as evidence of convective collapse. This case is similar to our observations as far as the formation of a bright point associated with a strong downflow is concerned, although our downflows do not reach supersonic speed. Notably, the flux tube retains a high field strength, unlike our cases. The modest brightening in the continuum at 630~nm for the Hinode data compared to the present SST data can probably be explained by the difference in spatial resolution, in agreement with \citet{2010A&A...509A..76D}. At the two times higher spatial resolution of the SST/CRISP data, we clearly see bright-point formation for all cases studied.
 
Downflows have also been observed in association with flux emergence \citep{2008A&A...481L..25I}. These authors found a strong downflows of $5.0$~\kmps at one of the foot points of the emerging loop.  \citet{2008ApJ...687.1373C} found supersonic downflows at flux cancellation sites in their flux emergence simulations. They also discussed Hinode observations of a supersonic downflow reaching $10.0$~\kmps exactly at the boundary of a flux cancellation region. We find weak opposite polarity field within our marker circles for cases $\bf{b}$, $\bf{g}$ and $\bf{h}$. However, we do not see evidence for supersonic downflows and the strongest downflows in all the three cases appear to be located well away from the expected reconnection site. Neither do we see any evidence of flux emergence related to these three cases. If indeed our observations are related to reconnection, we should also observe the (partial) disappearance of the magnetic field \citep{2010ApJ...712.1321K}. For our SST data, the involved flux concentrations are not sufficiently separated spatially from adjacent field to allow their integrated magnetic flux to be measured with confidence, but we can rule out the total disappearance of the magnetic field in all the three cases. Our interpretation of these three cases is that the weak opposite polarity fields are swept toward the stronger flux component by convective flows or that perhaps the weak opposite polarity field is formed as a result of overturning convection pulling some field lines into "serpentine" shape, as found in the simulations of  \citet{2010ApJ...720..233C}.

\section{Conclusion}
\label{sec:conclusion}
The high spatial and temporal resolution of our SST/CRISP spectropolarimetric data allows the detection of several small-scale  transient downflow events. The evolution of these events is studied by using Milne-Eddington inversions to extract magnetic field and LOS velocity at the locations of peak velocity. We demonstrate that the quality of the fits of the Stokes $I$ and $V$ profiles is improved dramatically by assuming spatial stray-light at a level that is consistent with the observed RMS granulation continuum contrast.

Our data are consistent with earlier observations and simulations describing flux tube collapse, but the transition to a state with stronger field appears transient and short-lived rather than resulting in a true transformation into a stable strong-field state. The present data shows no evidence of reconnection associated with the downflow events analyzed, presumably due to the lack of strong field flux emergence in this region at the time of the observations and the presence of only weak opposite polarity field within the FOV. Possibly, the weak opposite polarity field seen adjacent to three of our downflow sites can be explained by the effects of overturning convection pulling down some field lines and leading to small-scale up/down "serpentine" magnetic field.

\begin{acknowledgements}
The author is grateful to Andreas Lagg for valuable suggestions. The author would also like to thank G{\"o}ran Scharmer, Dan Kiselman and Mats L{\"o}fdahl for comments on the manuscript. The Swedish 1-m Solar Telescope is operated on the island of La Palma by the Institute for Solar Physics of the Royal Swedish Academy of Sciences in the Spanish Observatorio del Roque de los Muchachos of the Instituto de Astrof\'isica de Canarias.
\end{acknowledgements}

\begin{appendix}
\section{Cases b--h}
 \begin{figure*}[p]
   \includegraphics[bb=0 0 400 640,angle=90,width=\figwidth]{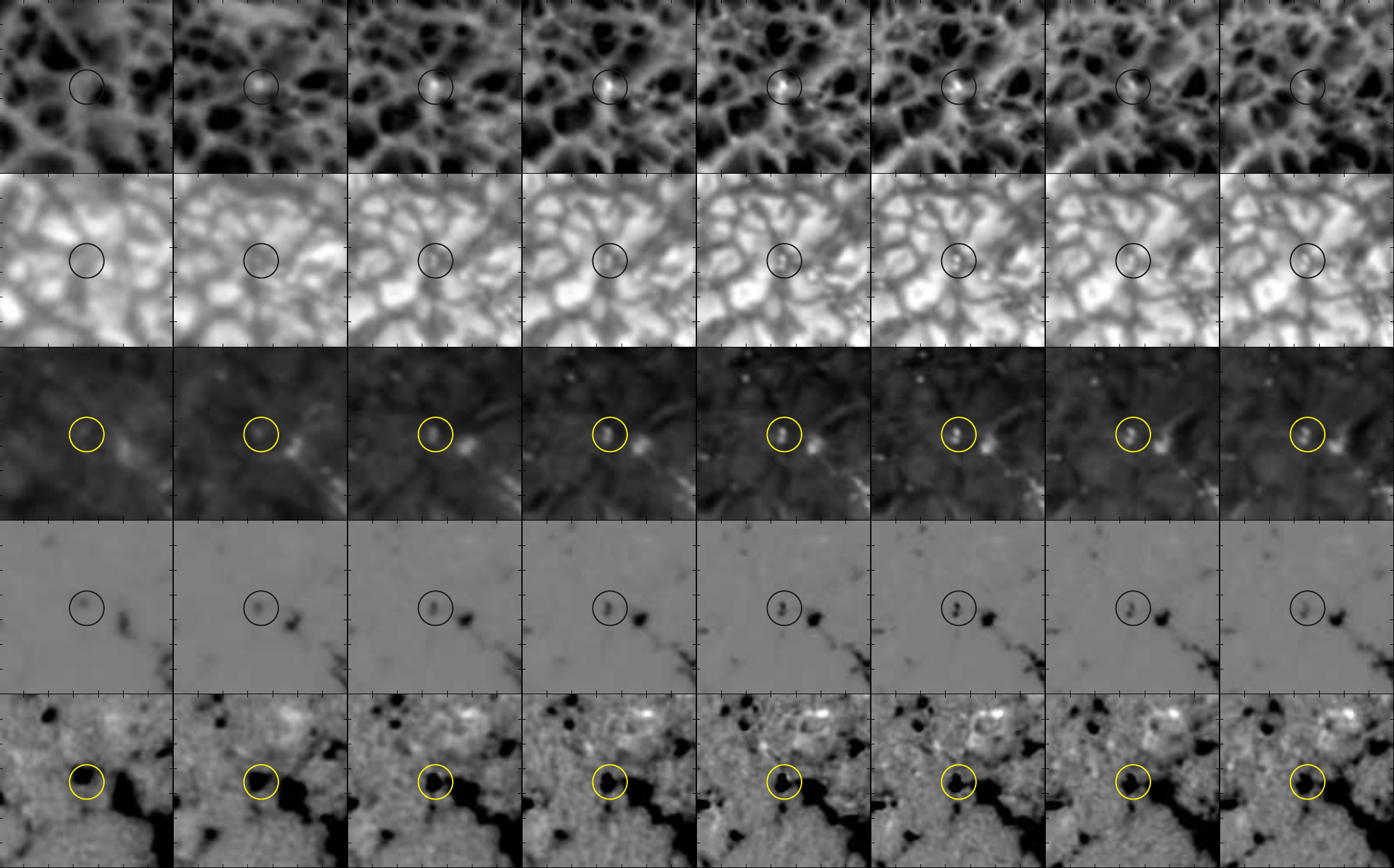}
\includegraphics[bb=0 0 87 634,width=0.077\textwidth]{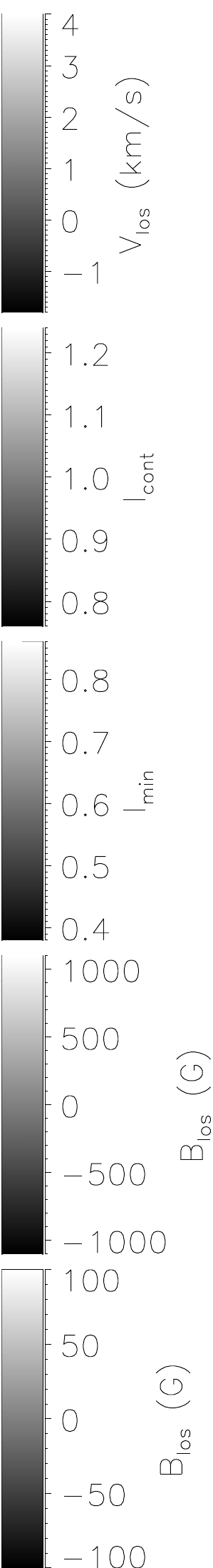}
 \caption{Case $\bf{b}$. See Fig. 5 for explanations}
    \label{}
  
\end{figure*}

\begin{figure*}[p]
    \includegraphics[bb=0 0 400 640,angle=90,width=\figwidth]{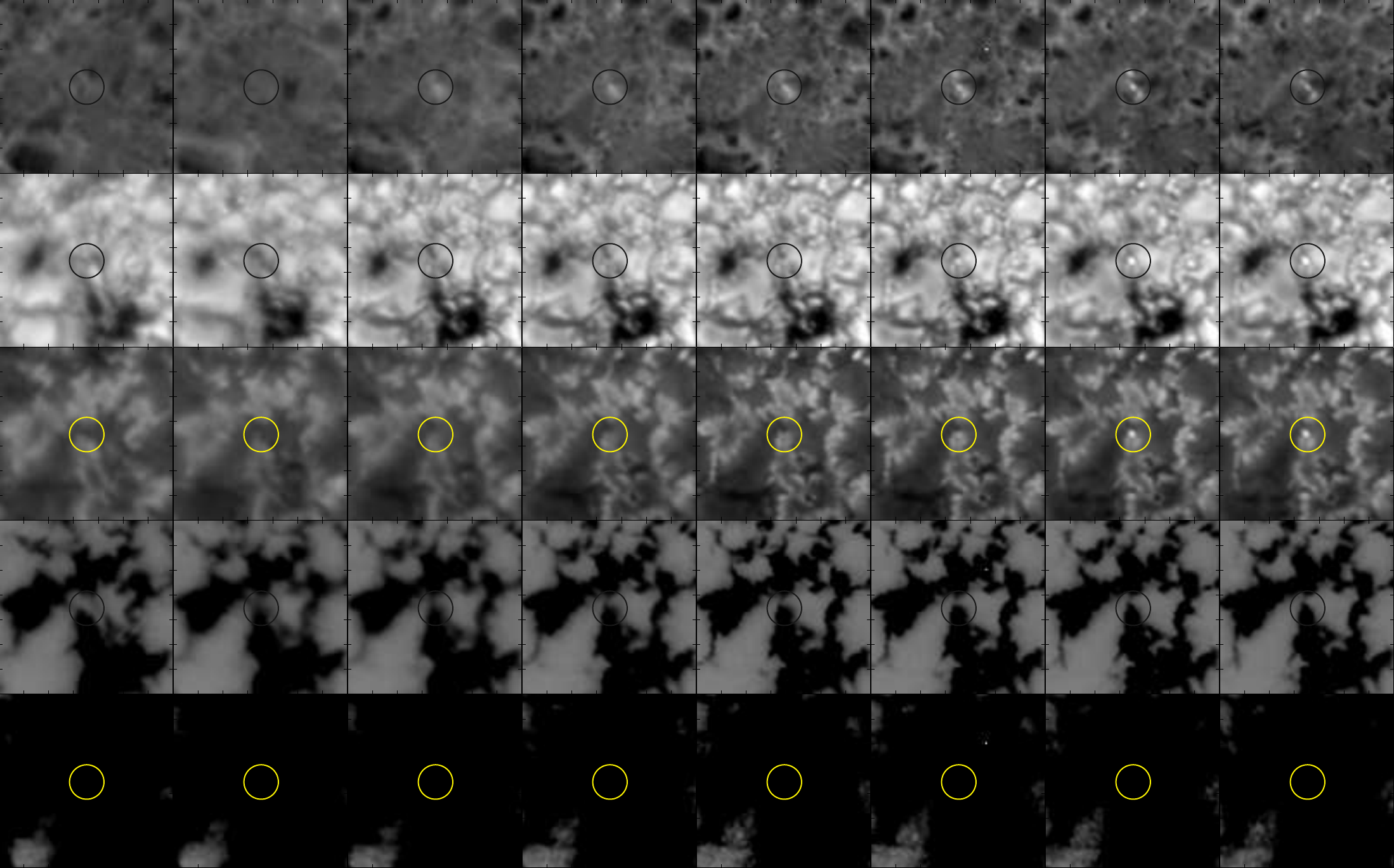}
\includegraphics[bb=0 0 87 634,width=0.077\textwidth]{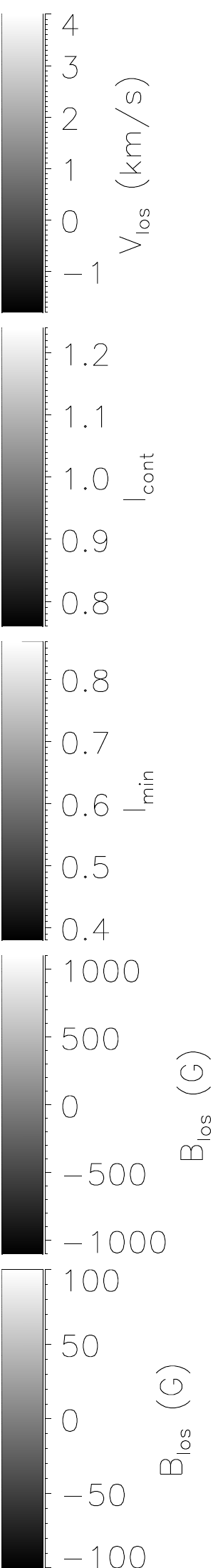}
 \caption{Case $\bf{c}$. See Fig. 5 for explanations}
    \label{}
  
\end{figure*}
\begin{figure*}[p]
   \includegraphics[bb=0 0 400 640,angle=90,width=\figwidth]{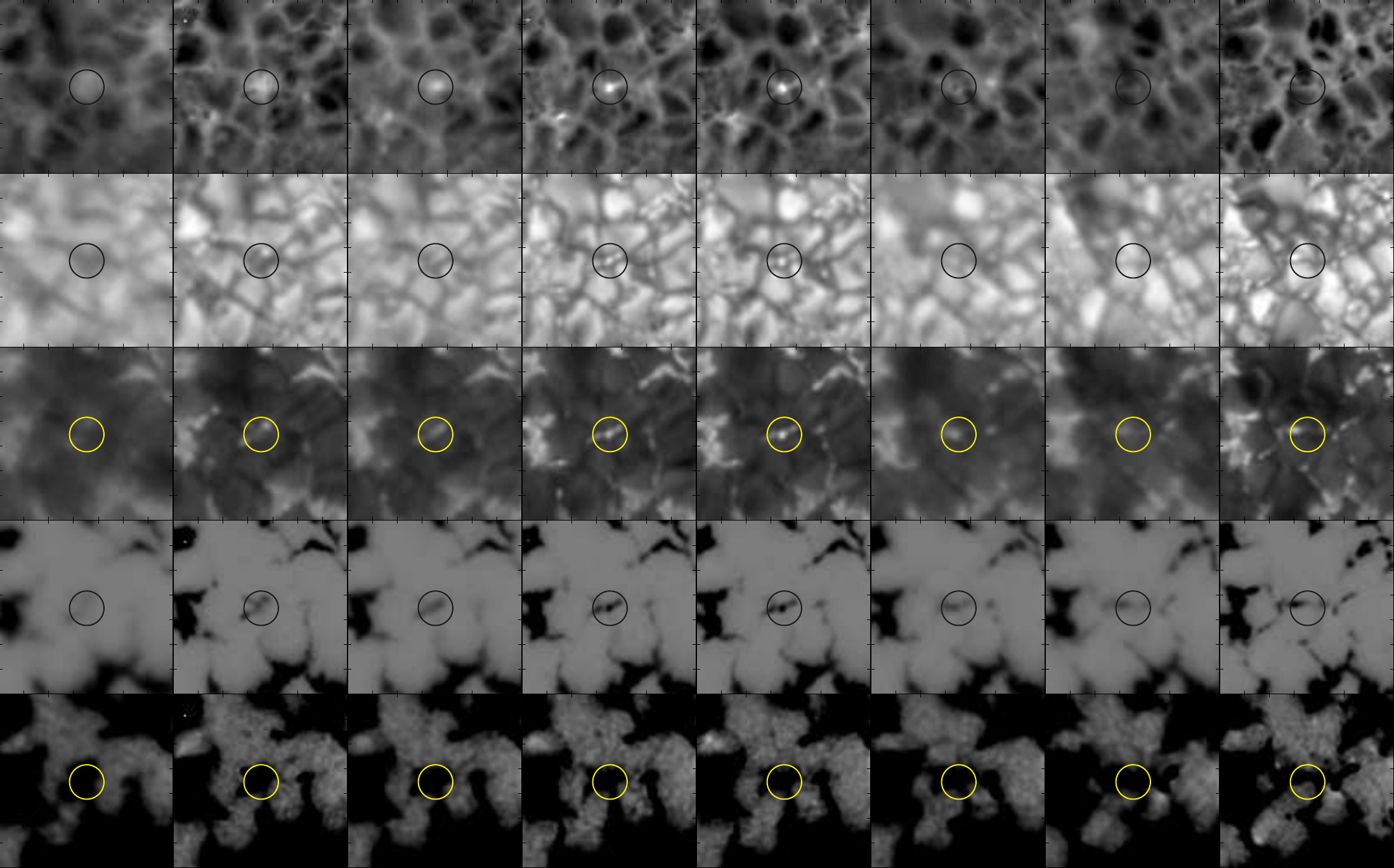}  
\includegraphics[bb=0 0 87 634,width=0.077\textwidth]{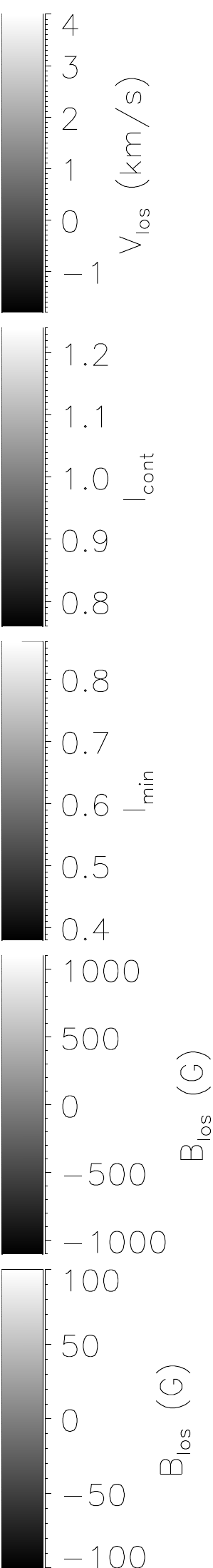}
 \caption{Case $\bf{d}$. See Fig. 5 for explanations. The snapshots shown are recorded at $t=0$, 84, 126, 189, 231, 315, 546, and 651 seconds (left to right).}
    \label{}
\end{figure*}

\begin{figure*}[p]
    \includegraphics[bb=0 0 400 640,angle=90,width=\figwidth]{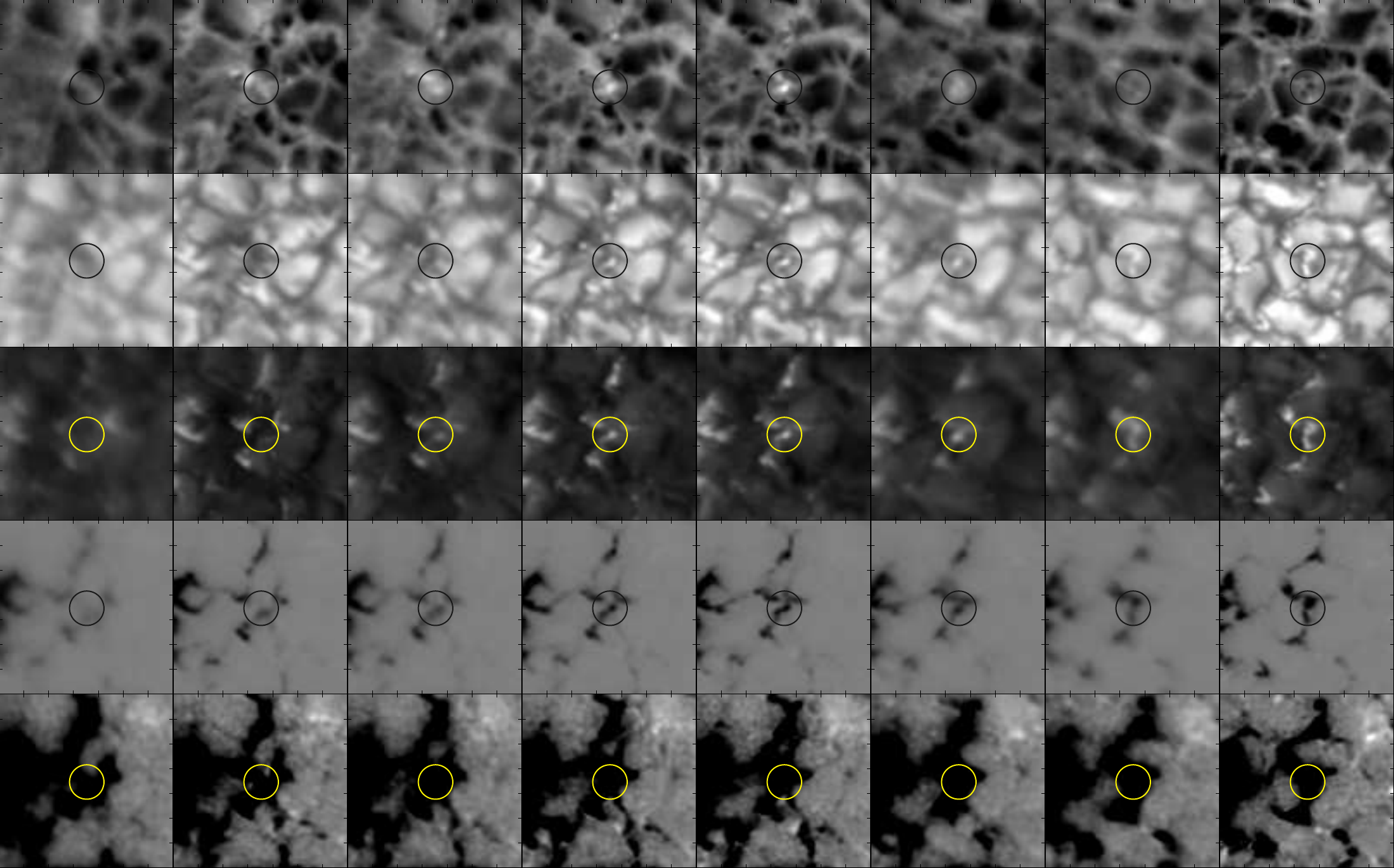}
\includegraphics[bb=0 0 87 634,width=0.077\textwidth]{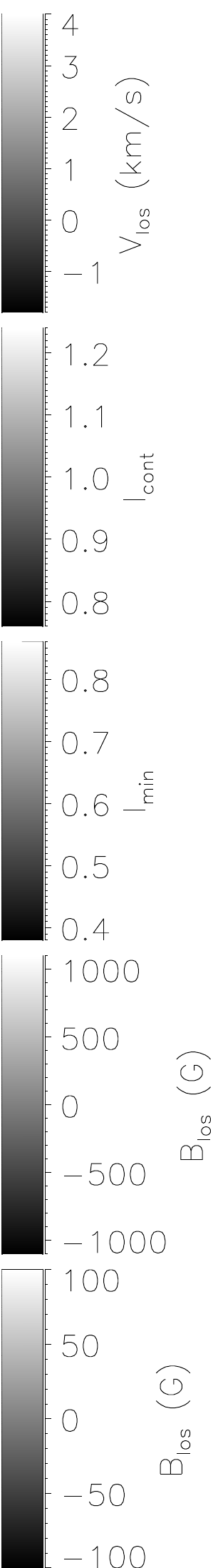}
 \caption{Case $\bf{e}$. See Fig. 5 for explanations. The snapshots shown are recorded at $t=0$, 84, 126, 189, 231, 315, 546, and 651 seconds (left to right).}
    \label{}
\end{figure*}
\begin{figure*}[p]
   \includegraphics[bb=0 0 400 640,angle=90,width=\figwidth]{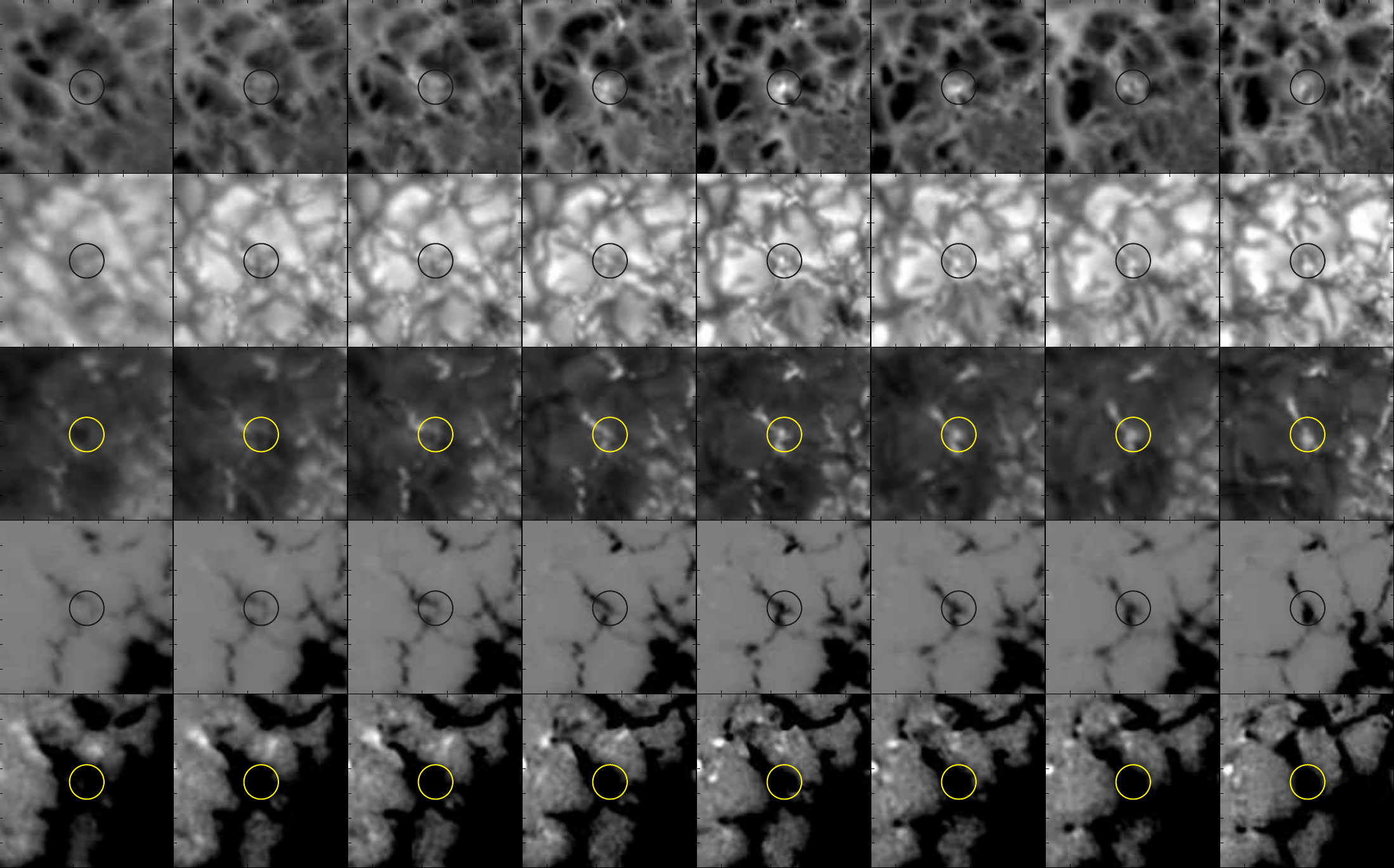}
\includegraphics[bb=0 0 87 634,width=0.077\textwidth]{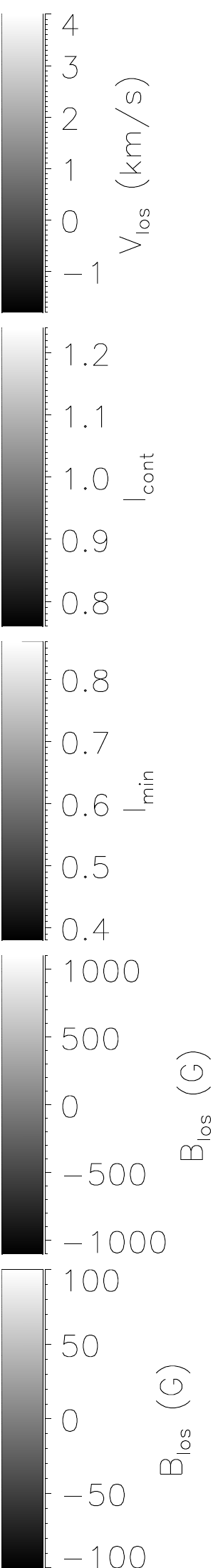}
 \caption{Case $\bf{f}$. See Fig. 5 for explanations. The snapshots shown are recorded at $t=0$, 84, 126, 210, 294, 336, 420, and 546 seconds (left to right).}
    \label{}
  
\end{figure*}
\begin{figure*}[p]
   \includegraphics[bb=0 0 400 640,angle=90,width=\figwidth]{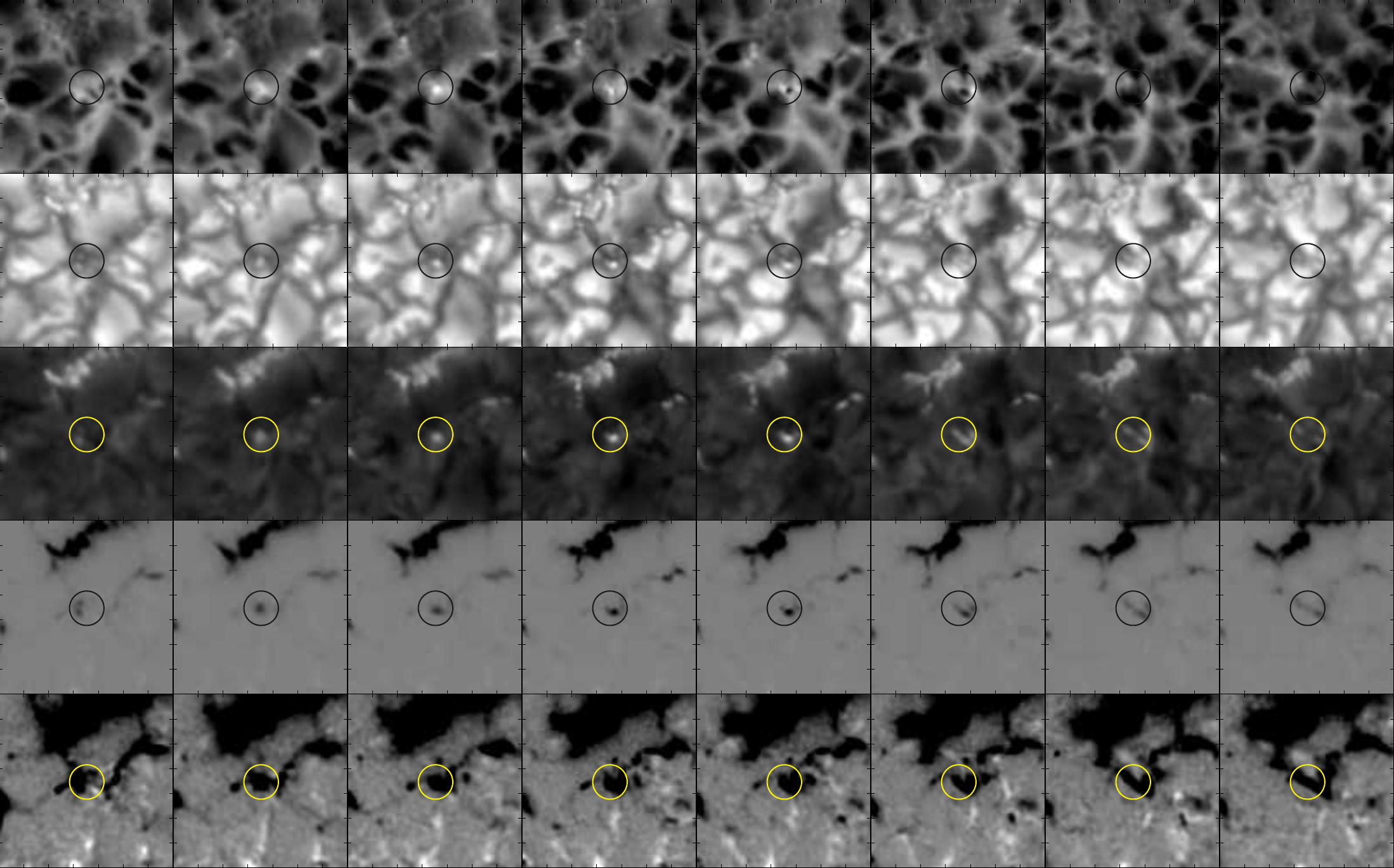}
\includegraphics[bb=0 0 87 634,width=0.077\textwidth]{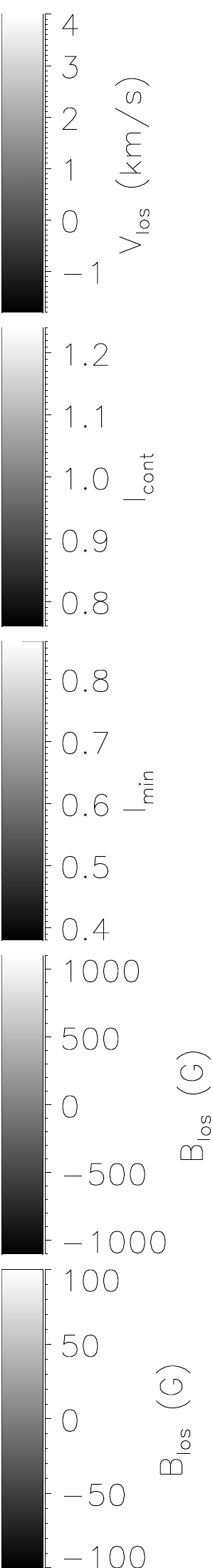}
 \caption{Case $\bf{g}$. See Fig. 5 for explanations. The snapshots shown are recorded at $t=21$, 126, 210, 294, 357, 441, 504, and 567 seconds (left to right). }
    \label{}
\end{figure*}

\begin{figure*}[p]
   \includegraphics[bb=0 0 400 640,angle=90,width=\figwidth]{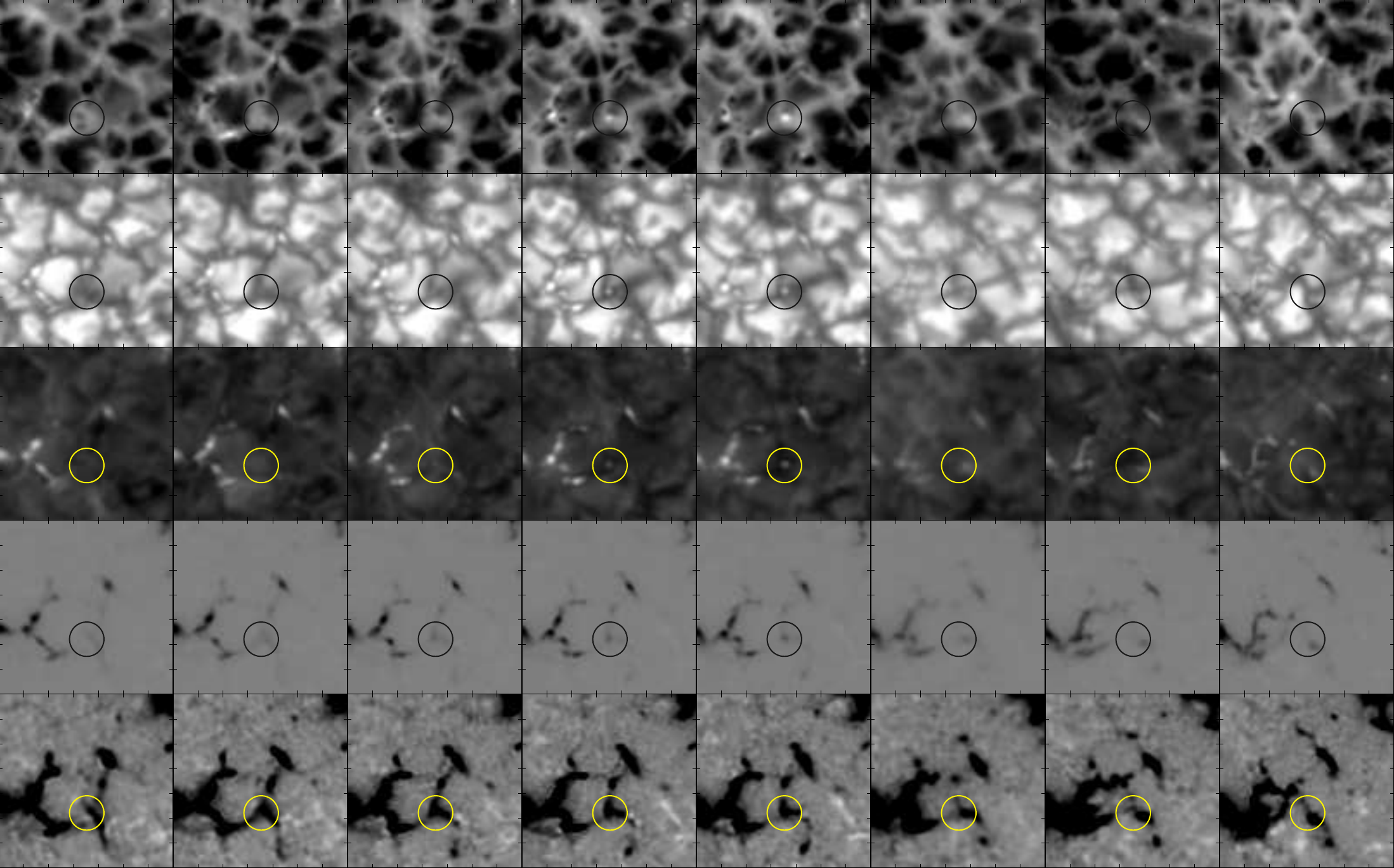}
\includegraphics[bb=0 0 87 634,width=0.077\textwidth]{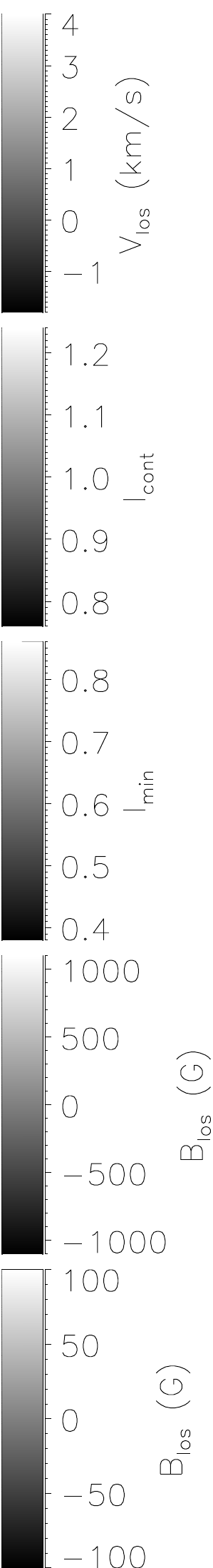}
\caption{Case $\bf{h}$. See Fig. 5 for explanations. The snapshots shown  are recorded at $t=21$, 105, 168, 210, 252, 357, 462, and 588 seconds (left to right). }
    \label{}
  
\end{figure*}
\end{appendix}


\end{document}